# A Bayesian-network-based cybersecurity adversarial risk analysis framework with numerical examples

Jiali Wang jiali.wang@qmul.ac.uk, Martin Neil m.neil@qmul.ac.uk

School of Electronic Engineering & Computer Science, Queen Mary University of London.

Corresponding author: Jiali Wang

Postal Address: School of Electronic Engineering & Computer Science, Queen Mary University of London, Mile End Road, London E1 4NS.

## Abstract

Cybersecurity risk analysis plays an essential role in supporting organizations make effective decision about how to manage and control cybersecurity risk. Cybersecurity risk is a function of the interplay between the defender (the organisation) and the attacker: decisions and actions made by the defender 'second guess' the decisions and actions taken by the attacker and vice versa. Insight into this 'game' between these two agents provides a means for the defender to identify and make optimal decisions. To date, the adversarial risk analysis framework has provided a decision-analytical approach to solve such game problems in the presence of uncertainty and uses Monte Carlo simulation to calculate and identify optimal decisions. We propose an alternative framework to construct and solve a serial of sequential Defend-Attack models, that incorporates the adversarial risk analysis approach, but uses a new class of influence diagrams algorithm, called hybrid Bayesian network inference, to identify optimal decision strategies. Compared to Monte Carlo simulation the proposed hybrid Bayesian network inference is more versatile because it provides an automated way to compute hybrid Defend-Attack models and extends their use to involve mixtures of continuous and discrete variables, of any kind. More importantly, the hybrid Bayesian network approach is novel in that it supports dynamic decision making whereby new real-time observations can update the Defend-Attack model in practice. We also extend the Defend-Attack model to support cases involving extra variables and longer decision sequence. Examples are presented, illustrating how the proposed framework can be adjusted for more complicated scenarios, including dynamic decision making.

# Contents

# 1. Introduction

Cybersecurity risk analysis plays an essential role in supporting organizations make effective decisions about how to manage and control cybersecurity risk. Cybersecurity risk is a function of the interplay between the defender (the organisation) and the attacker: decisions and actions made by the defender 'second guess' the decisions and actions taken by the attacker and vice versa. Insight into this 'game' between these two agents provides a means for the defender to identify and make optimal decisions.

Game-theoretical approaches have been the typical choice to model interplay between two or more strategic adversaries and have been widely applied to cybersecurity issues (Do et al., 2017; Manshaei et al., 2013; Roy et al., 2010; Wang et al., 2016). However, conventional game theory faces a challenge when it aims to find solutions for all the participants of the game, in that the solution must be a Nash equilibrium. As the problem and associated game models get more realistic and complex, this requirement makes it increasingly difficult to compute a solution (Joshi et al., 2020) (Gindis, 2009). Moreover, most versions of non-cooperative game theory assume adversaries know their own payoffs, preferences, beliefs and possible actions but also assume that knowledge about these is shared between adversaries (Harsanyi, 1967). This shared knowledge assumption is unrealistic in contexts such as counter-terrorism or cybersecurity, where players will not generally have sufficient knowledge about their opponents or where opponents are motivated to keep secrets (González-Ortega et al., 2019).

Adversarial Risk Analysis (ARA) (Rios Insua et al., 2009) was proposed to address the above mentioned shortcomings of classic game theory. ARA solves the problem by modelling the ability of a player (typically, the defender) to anticipate the opponent's decisions and actions. General security risk analysis problems, as explored in (Brown et al., 2006) (Zhuang & Bier, 2007) (Hausken & Bier, 2011), are modelled as a number of basic templates (i.e. simultaneous Defend-Attack (D-A) model, sequential D-A model, etc) with a known ARA solution in (Banks et al., 2015). The templates differ in the way and order in which the attack and defence actions take place within the global sequence of decisions and events, as well as in the information revealed. These templates can then be represented by Influence Diagrams (ID) (Fenton & Neil, 2019).

How to best model and efficiently calculate optimal decisions using ARA has received a lot of attention in recent years. Opponents in simultaneous decision making games are modelled following ARA in (Rios Insua et al., 2016). Insider threat in sequential D-A games were modelled using the ARA approach in (Joshi et al., 2020). A calculation procedure for conducting ARA for a bi-agent game is introduced in (González-Ortega et al., 2019). In the work (González-Ortega et al., 2019), a model consists of sequential D-A pattern and simultaneous D-A pattern is considered. For more practical cases, (Insua et al., 2019) provides an ARA framework for cybersecurity risk analyse using insurance as part of the security portfolio for decision making and the work done by (Gil & Parra-Arnau, 2019) applied ARA in Counterterrorist Online Surveillance.

It is argued in (Joshi et al., 2020), that in most realistic cybersecurity cases, the defender would deploy their defence first to deter and prevent attacks and, therefore, it makes sense to model the cybersecurity problem as a sequential D-A game, rather than as a simultaneous one. However, solving the sequential D-A model, and its more challenging extensions i.e., the sequential D-A model with extra variables or with a longer decision sequence, has not been systemically investigated in previous research. In this paper, we focus on solving the bi-agent sequential D-A game model and its extensions. We provide a Hybrid Bayesian Network (HBN) based ARA approach as a comprehensive solution and use examples to illustrate how the proposed framework can be applied to practical problems. Our proposed solution can be easily applied to solve other typical sequential game templates summarized in (Banks et al., 2015). For example, the D-A-D model can be regarded as an instance of the sequential D-A model with longer decision sequences. Moreover, solving sequential A-D models (Banks et al., 2015) and the extensions (i.e., sequential A-D-A models) can be regarded as a reflective solution to the "dual problem" of solving the D-A problem, since only the order of making decision changes while the underlying calculating mechanism remains the same compared with solving the D-A model.

Most ARA solutions use Monte Carlo (MC) simulation to carry out the calculation, for example in (Insua et al., 2019) (Rios Insua et al., 2016) (Joshi et al., 2020) (González-Ortega et al., 2019). MC simulation is straightforward to implement. However, this approach can become computationally challenging when dealing with decision dependent uncertainties, especially in D-A models where we encounter longer decision sequences. Moreover, it cannot cope with new evidence that could be used to update the game model, dynamically, in real time, which we contend is a realistic requirement for practical use. In our work, we provide algorithms to implement the ARA approach based on the HBN inference (Yet et al., 2018). The proposed method offers a fully automated way to compute sequential D-A models and can support dynamic decision analysis which has not been solved by previous ARA solutions that adopt MC simulation.

The contributions of our work are threefold: 1) we propose an alternative framework, based on HBN inference and decision trees, to solve the typical sequential Defend-Attack (D-A) games from the ARA perspective; 2) we illustrate how to use this framework to solve more practical D-A problems involving extra variables and longer decision sequence (also known as multi-period game/ k-level thinking); 3) we present numerical examples to show how our framework can support decision making in different application contexts involving extra variables, longer decision sequences and dynamic decision making (in section 4 and 5). The advantages of the proposed framework compared with other previous works are: 1) it offers a fully automated way to compute hybrid D-A models which involve mixtures of continuous and discrete variables; 2) it can provide more flexible decision making about risk in addition to expected utility optimization; 3) it supports dynamic decision making in multi-period D-A games.

The rest of the paper is structured as follows. In section 2, we introduce technologies adopted in our work, including hybrid Bayesian networks, influence diagrams and decision trees. In Section 3 we show how we can implement the ARA approach incorporating HBNs using a typical game model: the sequential D-A model. In section 4, we summarize the rules needed to be followed to extend D-A models for more complicated scenarios, i.e., D-A problems with extra variables and with longer decision sequences and apply these rules to two examples. In section 5, we discuss dynamic decision analysis, provide the algorithm and illustrative examples. Conclusions are presented in section 6.

## 2. Influence Diagrams, Hybrid Bayesian Networks and Decision Trees

In ARA, a decision problem can be structured and represented by an Influence Diagram (ID), which is a generalization of a Bayesian network. In this section, we provide a general introduction of ID, Hybrid Bayesian Networks (HBNs) and how to conduct decision analysis through HBN using Decision Trees (DTs).

BNs are widely used for probabilistic reasoning and have very wide applicability, including enabling statistical reasoning, causal reasoning and diagnostic inference (Fenton & Neil, 2011). A BN is a directed acyclic graph representing a factorization of a joint probability distribution, consisting of nodes representing variables and arcs representing causal or probabilistic relationships (the qualitative part) with probabilistic weights (the quantitative part) sometimes modelled using statistical and deterministic functions. In a BN, each node $X_i$ has an associated probability table, $P\big((X_i | pa(X_i)\big)$, called the Conditional Probability Table (CPT) of $X_i$ given its parent variables, $pa(X_i)$. For a node $X_i$ without parents, the CPT is the marginal probability distribution of $X_i$, $P(X_i)$. The conditional-independent relationship among variables, represented by the absence of arcs, allows simplification of a BN's joint probability distribution which can be represented by the product of CPTs. Furthermore, the marginal distribution of the child variable can be obtained by marginalizing over its parent variables in the joint distribution (Fenton & Neil, 2019). For example, considering a simple BN consisting of three nodes, in which nodes $A$ and $B$ are parents of node $C$, and CPTs are $P(A)$, $P(B)$ and $P(C|A, B)$, we can get the joint distribution of this BN from $P(A, B, C) = P(A) P(B) P(C|A, B)$ and calculate the marginal distribution of the child node $C$ following $P(C) = \sum_{A,B} P(A, B, C)$.

Generally, the joint distribution of a BN can be calculated following formula (1):

$$P(X_1, \dots, X_n) = \prod_{i=1}^{n} P\big((X_i | pa(X_i)\big) \tag{1}$$

This factorisation significantly reduces the complexity of inference tasks in BNs.

Computational algorithms for solving BNs, where all variables are discrete, include Junction Tree (JT) (Lin et al., 2014). Solutions for continuous cases include Hybrid BNs Junction Tree algorithm which can be combined with Dynamic Discretization (DD) (Fenton & Neil, 2019), to calculate BN containing mixtures of continuous and discrete variables of any distributional form. HBNs have been implemented and packaged into the BN modelling software application AgenaRisk (*Agena Ltd. 2002-2021*). In this paper we have used AgenaRisk, a commercial BN package, to build BNs and perform calculations for ARA. AgenaRisk contains off-the-shelf functions for performing inference on HBNs and influence diagrams.

An Influence Diagram (ID) is a special type of BN models that represents the interaction between decisions, chance variables, and utilities along with an algorithm to compute the expected utilities and identify those decisions that optimise the utility (Fenton & Neil, 2019) (Yet et al., 2018). In an ID, nodes represent each variable, with the convention that rectangles represent decisions, ovals represent chance variables, and diamonds represent utilities. Each decision node represents a decision, chance nodes represent random variables, which can be observable or non-observable, and utility nodes represent the pay-off for the decision maker. There is always an "ultimate" utility node (with at least one parent) that we seek to optimize. An ID with chance, decision, and utility nodes that can involve mixtures of discrete and continuous variables are called Hybrid IDs (HIDs) which is a kind of Hybrid BNs (HBNs).

An ID of the sequential D-A model is shown in Figure 1. This ID represents a sequential D-A game where the defender would make her defence deployment decision (represented by node $D$) first. This can then be observed by the attacker (i.e., he) and he would use this information to optimise his attack decision, i.e., whether to attack or use how much resource for attacking (represented by node $A$). Whether the attack is successful is represented by the chance node, $S$, which is conditional on $D$ and $A$. Finally, $D$ and $S$ determine $D's$ utility while $A$ and $S$ determine $A's$ utility. The decision that maximizes $D's$ utility would be the optimal strategy for the defender. The rule for the attacker is similar.

Figure 1  Influence diagram for the sequential Defend-Attack game

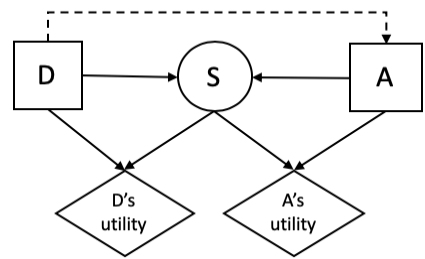

Generally, in an ID, incoming arcs to chance or utility nodes represent causal, deterministic, or associational relations between the node and its parents. Incoming arcs to decision nodes (shown by a dashed line) are "informational" arcs, representing the possibility that the temporal ordering, when the state of any parent node might be known before a decision is made. Informational (dashed line) arcs also specify the sequential order of decisions and observations. An ID cannot be computed without a strict sequential order.

After constructing an ID for a decision analysis problem, we can construct a Decision Tree (DT) (Fenton & Neil, 2019) to represent all the potential decisions and their corresponding utility values. The decision which corresponds to the maximum (in general) utility would be determined to be the optima in the decision tree. We construct DTs for an ID using a hybrid ID (Yet et al., 2018) in AgenaRisk.

Adversarial Risk Analysis (ARA) provides a decision analytic approach offering prescriptive supporting one of the intervening agents (i.e., the defender) based on an expected utility model treating the adversary's decisions as uncertainties. As we have mentioned, since it is rational to model the cybersecurity problem as a sequential D-A game, rather than as a simultaneous one (Joshi et al., 2020), we focus on solving the sequential D-A model and its extensions in this paper. Fundamentally, ARA solves the D-A game by analysing the attacker's problem, anticipating his best choice and taking into account the defender's own options for the most optimal defence strategy. In section 3, we formally illustrate how to use the HBN to solve the typical D-A game from the ARA perspective.

# 3. The sequential Defend-Attack model

## 3.1 Adversarial risk analysis of the Defend-Attack model

The adversarial risk analysis of the sequential Defend-Attack (D-A) game model provides a template and procedure to identify the optimal strategy for the defender. In this section, we analyse the D-A game represented by Figure 1 from the ARA perspective and illustrate how to construct an HBN for the calculation and supporting decision making for the defender.

We assume that in the ID represented by Figure 1, the defender has a discrete set of possible defence levels, which are represented by the decision node $D$ (Defences). After observing the potential defence levels that can be implemented, the attacker creates a discrete set of possible attack levels $A = \{a_1, a_2, ..., a_m\}$ represented by node $A$ (Attacks). A dashed arc pointing from node $D$ to $A$ represents the fact that the attacker's decision depends on the potential defence. From the ARA perspective, the D-A game can be divided into the attacker's problem and the defender's problem.

To determine the defender's best choice, the defender would analyse the attacker's problem first, which is represented by Figure 2, to anticipate choices the attacker might optimally make. Then, based on this analysis, the defender would determine the optimal strategy for herself in the first place. This calculation procedure is an implementation of backwards induction (Aliprantis & Chakrabarti, 2012) (Banks et al., 2015). Backwards induction analyses decisions from the end to the beginning of the decision sequence to calculate optimal strategies for decision nodes. Assuming rationality, the attacker should choose the strategy that can maximize his utility, given all the potential defence choices, and that the defender will take this into account.

Figure 2  The attacker's problem in the D-A game

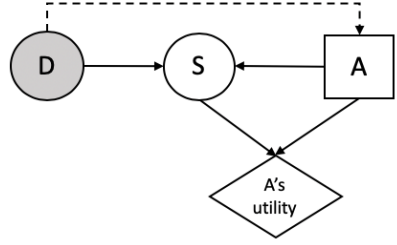

$A's$ expected utility corresponding to each possible combination of $(d,a) \in D \times A$ is:

$$\Psi_A(d,a) = p_A(S=0|d,a)u_A(d,a,S=0) + p_A(S=1|d,a)u_A(d,a,S=1) \qquad (2)$$

Therefore, the defender can predict that the optimal attack that would be adopted is:

$$a^*(d) = argmax_{a \in A}\Psi_A(d,a), \forall d \in D \qquad (3)$$

Consequently, the defender can calculate her optimal initial strategy to adopt in the game by analysing the defender's problem which is shown in Figure 3.

Figure 3  The defender's problem in the D-A game

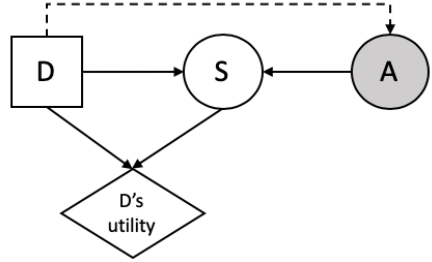

The expected utility of $D$ corresponding to each possible combination of $(d,a) \in D \times A$ is:

$$\Psi_D(d,a) = p_D(S=0|d,a)u_D(d,a,S=0) + p_D(S=1|d,a)u_D(d,a,S=1) \qquad (4)$$

Under the assumption that the opponent in this game is rational, her best choice is:

$$d^* = argmax_{d \in D}\Psi_D\big(d,a^*(d)\big) \qquad (5)$$

This calculation follows backwards induction as is represented that, the decision sequence in reality is from $D$ to $A$, while the analysing/calculating sequence is backwards, from $A$ to $D$.

Note that, in contrast with classic game theory, the solution $d^*$ for the sequential game need not correspond to a Nash equilibrium, since in ARA, players are not assumed to have full and common knowledge and the solution $d^*$ is derived from the predicted $(p_A, u_A)$ rather than the actual one (Banks et al., 2015).

## 3.2 Depicting the Defend-Attack game problem using hybrid Bayesian networks

Here we use an example to show how to implement the sequential D-A game using a Hybrid Bayesian Network (HBN), that models a concrete sequential Defend-Attack game, as shown in Figure 4 (a). We simplify the opponents' decisions as Boolean variables representing to defend or not, for the defender, and to attack or not, for the attacker. We assume that the defender's decision is about whether to defend an information asset. Meanwhile, after observing whether the defender defends, the attacker would consider whether to attack. We show the setting of nodes' CPTs in Figure 4 (b).

We assign uniform distributions to the decision nodes: node $D$ (defence decision) and node $A$ (attack decision) representing the opponents' open-mindedness choices. The CPT of the Success node (node $S$) models how the attack and defence interact to determine the probability of a successful attack. The node $S$ can be true or false. If an attack is not made, the probability of $S$ to be true is zero. We assume that if an attack is made, the probability of success is 0.8 if the asset is undefended, while it decreases to 0.2 if defended. The utility node, $D's$ Utility, models the defender's payoff given the asset is defended (utility: -100) and the cost to the defender of a successful attack is (utility: -200). The defender can predict the attacker's utility based on the assumption that the attacker's attack cost (utility: -100) and the payoff from a successful attack (utility: +200).

Figure 4  The BN model of a sequential D-A game

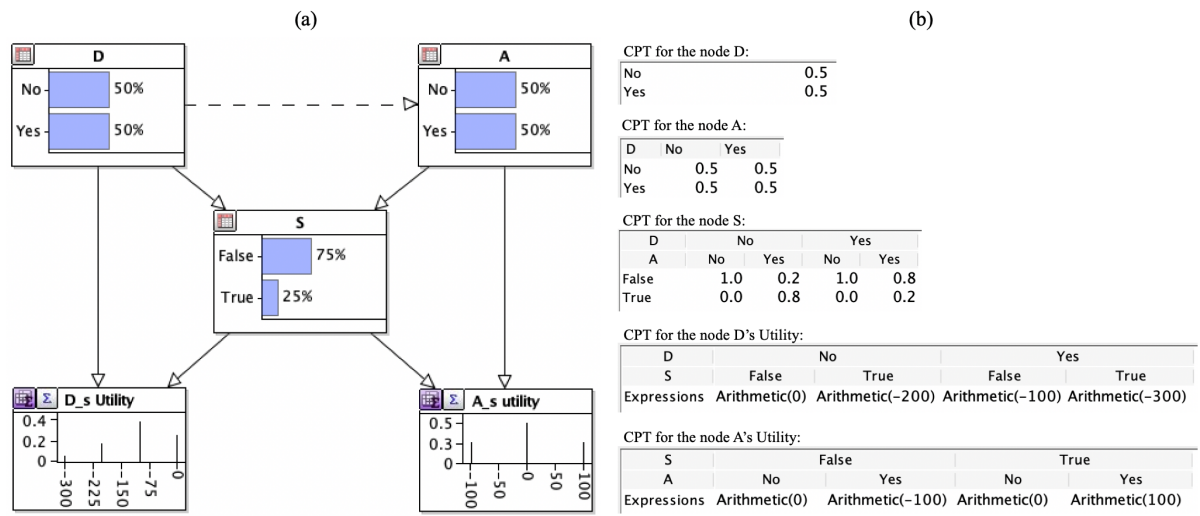

## 3.3 Risk assessment and decision support for the defender

In subsection 3.1, backwards induction is introduced to determine the optimal strategy for the defender in the D-A model in general. In this subsection, we illustrate how to implement the backwards induction for calculating the optimal decision for the defender in the HBN shown in Figure 4. To achieve this, there are three steps involved.

Firstly, the defender would initially analyse the attacker's problem, as shown in Figure 2, to predict what attacks he might make against possible defences. At this point, we regard the defender decision choices as a variable that might be potentially observed by the attacker and used to inform his decision making. Since the decision node for the defender becomes a chance node in this subproblem, we use an oval node to represent it. Here, the attacker's judgment is that there is a fifty-fifty chance of the defender defending or not. We calculate the attacker's utility of attacking, or not, under the two scenarios and identify those choices that maximize his utility, given he observes the defender's action. This calculation follows formula (2). In the Influence Diagram Analysis Function in AgenaRisk, this calculation can be done automatically by selecting node $A$ to be the decision node, node $D$ and $S$ to be the chance node and node $A's$ Utility to be the utility node (*Agena Ltd. 2002-2021*). In this example, we aim to maximize the expected utility and the calculation results are graphically represented by the Decision Tree (DT) shown in Figure 5.

Figure 5  The DT of the attacker's problem

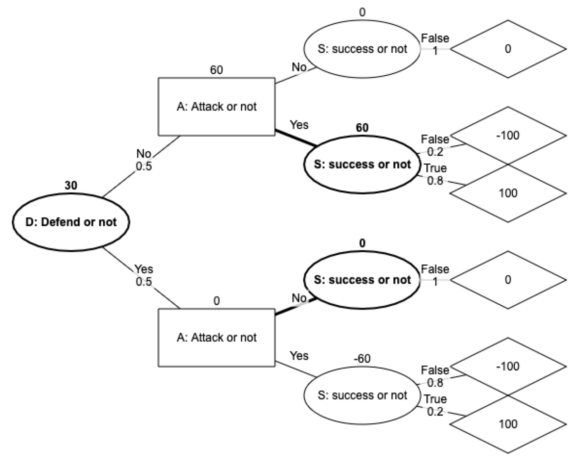

Figure 5 shows the best decision for the attacker in bold arcs, occur when the defender does not defend herself, and the best choice for the attack is to attack, which provides him the maximum utility (60), while if the defender defends, the attacker's best choice would be to not attack, with the maximum utility (0).

The second step is to update the CPT of the attacker's decision, $A$, in the model. The CPT for node $A$ is not 0.5 vs 0.5 anymore. The updated CPT for $A$ is $P(A = No|D = Yes) = 1$, $P(A = Yes|D = Yes) = 0$, $P(A = No|D = No) = 0$ and $P(A = Yes |D = No) = 1$ according to the results represented by Figure 5. We illustrate the updated D-A model in Figure 6.

Figure 6  The updated D-A model

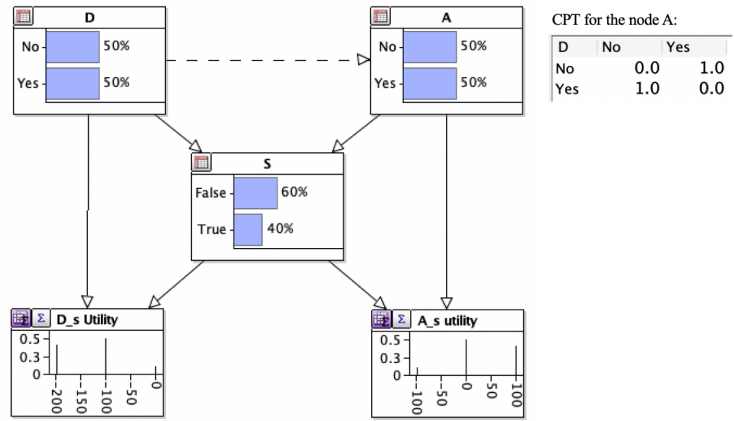

Then, in the last step, we determine the defender's optimal decision by analysing the defender's problem (shown in Figure 3) using the updated D-A model. In this step, the decision of the attacker is regarded as a variable dependent on the defender's decision. Hence, node $A$ becomes a chance node in the defender's problem. We calculate the defender's utility of defending or not following formula (4). We choose node $D$ as the decision node, node $A$ and $S$ to be the chance nodes and node $D's$ Utility as the utility node. The calculation results are graphically represented by the DT shown in Figure 7.

Figure 7  The DT of the defender's problem

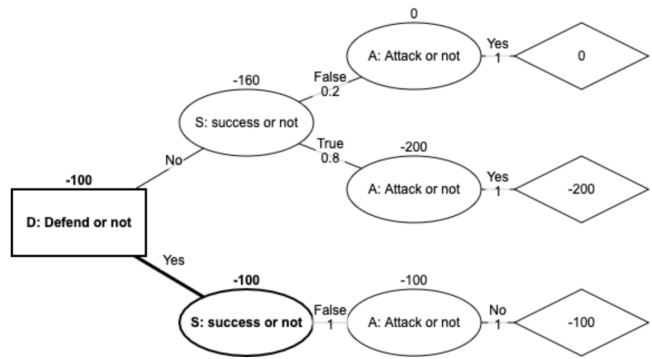

Here the optimal choice for the defender is shown by the bold arc, where to maximize her utility (-100) she should defend, otherwise she would suffer from a worse expected payoff (-160) if she does not. Therefore, the optimal decision for the defender is $D = Yes$ and sequentially, the attacker is anticipated that might not conduct the attack ($A = No$).

# 4. The sequential Defend-Attack-Defend model

## 4.1 Adversarial risk analysis of the Defend-Attack-Defend model

The sequential Defend-Attack-Defend (D-A-D) game model is an extension of the sequential Defend-Attack game. It assumes that the Defender (she) moves first (at time t = 1), making her initial defence deployment $D_1$. The Attacker (he) observes this choice and responds $A_2$ at time t = 2. Finally, the Defender chooses an action $D_3$ to mitigate the damage from the attack at time t = 3. Chance nodes $S_2$ and $S_3$ indicate the random payoffs that the defender and attacker receive on the second and third days of the game. It is assumed that the defender's utility consists of the cost of $D_3$ and the outcomes from $S_2$ and $S_3$ (Banks et al., 2015). Figure 8 shows the Influence Diagram (ID) of the D-A-D game problem. Previous work on this game has been done by (Lysyanskaya & Triandopoulos, 2006) (Rios & Insua, 2012) (Shan & Zhuang, 2013).

Figure 8  Influence diagram for the sequential Defend-Attack-Defend game

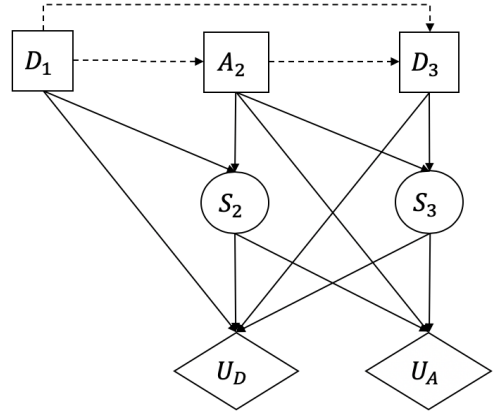

To determine the best choice for the defender, backward induction is implemented, which follows the similar idea with the procedure in the D-A model. The backward induction in the D-A-D model has five steps:

1) searching the optimal strategy of the subgame $D_3$, meaning determining choices in $D_3$ that can maximize the defender's expected utility, corresponding to each possible pair of $D_1$ and $A_2$; the expected utility function of the defender is determined by formula (6):

$$\Psi_D(d_{3i}, d_1, a_2) = P_D(S_3 = 0 | d_{3i}, d_1, a_2) \times U_D(d_{3i}, d_1, a_2, S_3 = 0) + \\ P_D(S_3 = 1 | d_{3i}, d_1, a_2) \times U_D(d_{3i}, d_1, a_2, S_3 = 1)$$ (6)

The optimal $D_3$ is determined by formula (7):

$$d_3^* = argmax_{d_3 \in D_3} \psi_D(D_1, A_2, d_3)$$ (7)

2) using results generated from step 1 to update the CPT of $D_3$ from the model.

3) analyse the attacker's problem assuming he can observe the defender's action $D_1$ and can predict her decision $D_3$. The expected utility function of the attacker is determined by formula (8)

$$\Psi_A(a_{2i}, d_1, d_3^*) = P_A(S_3 = 0 | a_{2i}, d_1, d_3^*) \times U_A(a_{2i}, d_1, d_3^*, S_3 = 0) + \\ P_A(S_3 = 1 | a_{2i}, d_1, d_3^*) \times U_A(a_{2i}, d_1, d_3^*, S_3 = 1)$$

(8)

The optimal $A_2$ is determined by formula (9):

$$a_2^* = argmax_{a_a \in A_2} \Psi_A(D_1, a_2, d_3^*)$$

(9)

4) using results from step 3 to update the CPT for $A_2$ from the model.

5) searching the best choice for $D_1$ that can maximize the defender's expected utility. That is defined by formula (10):

$$\psi_D(d_{1i}, d_3^*, a_2^*) = P_D(S_2 = 0 | d_{1i}, d_3^*, a_2^*) \times U_D(d_{1i}, d_3^*, a_2^*, S_2 = 0) + \\ P_D(S_2 = 1 | d_{1i}, d_3^*, a_2^*) \times U_D(d_{1i}, d_3^*, a_2^*, S_2 = 1)$$

(10)

The optimal $D_1$ is determined by formula (11):

$$d_1^* = argmax_{d_1 \in D_1} \psi_D(d_1, a_2^*, d_3^*)$$

(11)

Finally, the calculated results $D_1 = d_1^*$ and $D_3 = d_3^*(d_1^*, a_2^*)$ are optimal decisions for the defender while the attacker is predicted to adopt $A_2 = a_2^*(d_1^*)$ as his optimal action.

Figure 9 The defender's problem in the D-A-D game

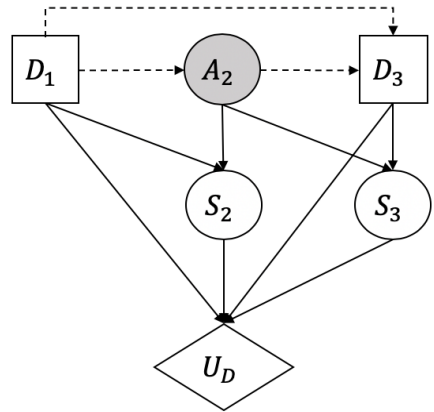

Figure 10  The attacker's problem in the D-A-D game

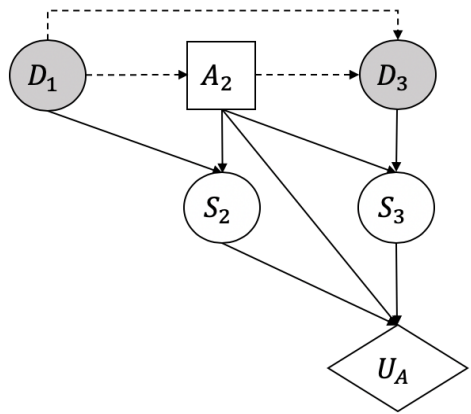

## 4.2 Depicting a D-A-D game problem using an HBN

In this subsection, we use an example to show how to implement a sequential D-A-D game using an HBN, evaluate involved cyber risks and support the defender to make the optimal defence decision.

We build an HBN to model a practical sequential D-A-D game as shown in Figure 11 (a). The defender and the attacker take actions alternatively towards an information system. In this model we use uniformly distributed variables with integer intervals to represent defence levels that the defender would decide to deploy and the attack levels that the attacker would decide to adopt. We assume that the defender's decision at time t = 1 is to equip a defence from level 0 to 3, to protect a target information asset. Level 0 means no defence is deployed. After observing the defender's deployment, the attacker would consider which attack level to adopt at time t = 2.  Finally, the defender chooses a defence level to mitigate the damage from the attack at time t = 3. The CPT of Success nodes (node S2 and S3) models how the attack and defence interact to determine the probability of a success attack at time t = 2 and time t = 3. They can be determined using comparative expressions in AgenaRisk. We assume only when the attack level is larger than the defence level, the attack would success. We set the CPT for both S2 and S3 following this idea. The utility node, D_s Utility, models the defender's payoff given the defence is deployed (utility: -50 multiplies the adopted defence level) and the cost of being attacked successfully (utility: -100 for one successful attack). The utility node (A_s Utility) models the attacker's payoff of conducting an attack (utility: -50 multiplies the adopted attack level) and the gain of a successful attack (utility: 100 for one successful attack). We represent CPT settings of involved nodes in Figure 11 (b).

Figure 11  The BN modelling a sequential D-A-D game

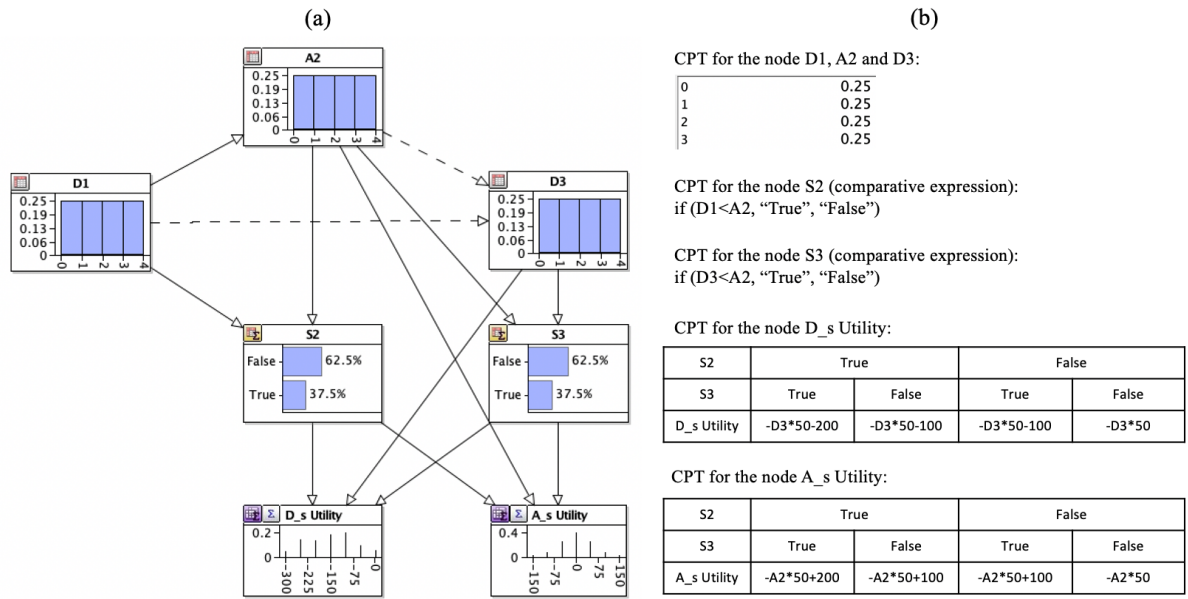

CPT for the node D1, A2 and D3:

| 0 | 0.25 |
|---|------|
| 1 | 0.25 |
| 2 | 0.25 |
| 3 | 0.25 |

CPT for the node S2 (comparative expression):
if (D1<A2, "True", "False")

CPT for the node S3 (comparative expression):
if (D3<A2, "True", "False")

CPT for the node D_s Utility:

| S2 | True | | False | |
|---|---|---|---|---|
| S3 | True | False | True | False |
| D_s Utility | -D3*50-200 | -D3*50-100 | -D3*50-100 | -D3*50 |

CPT for the node A_s Utility:

| S2 | True | | False | |
|---|---|---|---|---|
| S3 | True | False | True | False |
| A_s Utility | -A2*50+200 | -A2*50+100 | -A2*50+100 | -A2*50 |

## 4.3 Risk assessment and decision support for the defender

To determine the best choice for the defender, we conduct backward induction following procedure introduced in the subsection 4.1. Firstly, the defender would consider the defender's problem shown in Figure 9, that corresponding to each possible combination of $D_1$ and $A_2$ what would be the best defence strategy at time t = 3. The defender's expected utilities with respect to $D_3$, given $D_1$ and $A_2$, can be calculated following formula (6). In the Influence Diagram Analysis Function in AgenaRisk, this calculation can be done automatically by selecting node $D_3$ to be the decision node, nodes $D_1$ and $A_2$ to be the chance nodes and the node D_s Utility to be the utility node. The calculation results are graphically represented by a Decision Tree (DT) shown in Figure 12.

Figure 12  The DT with $D_3$ being the decision node in the defender's problem

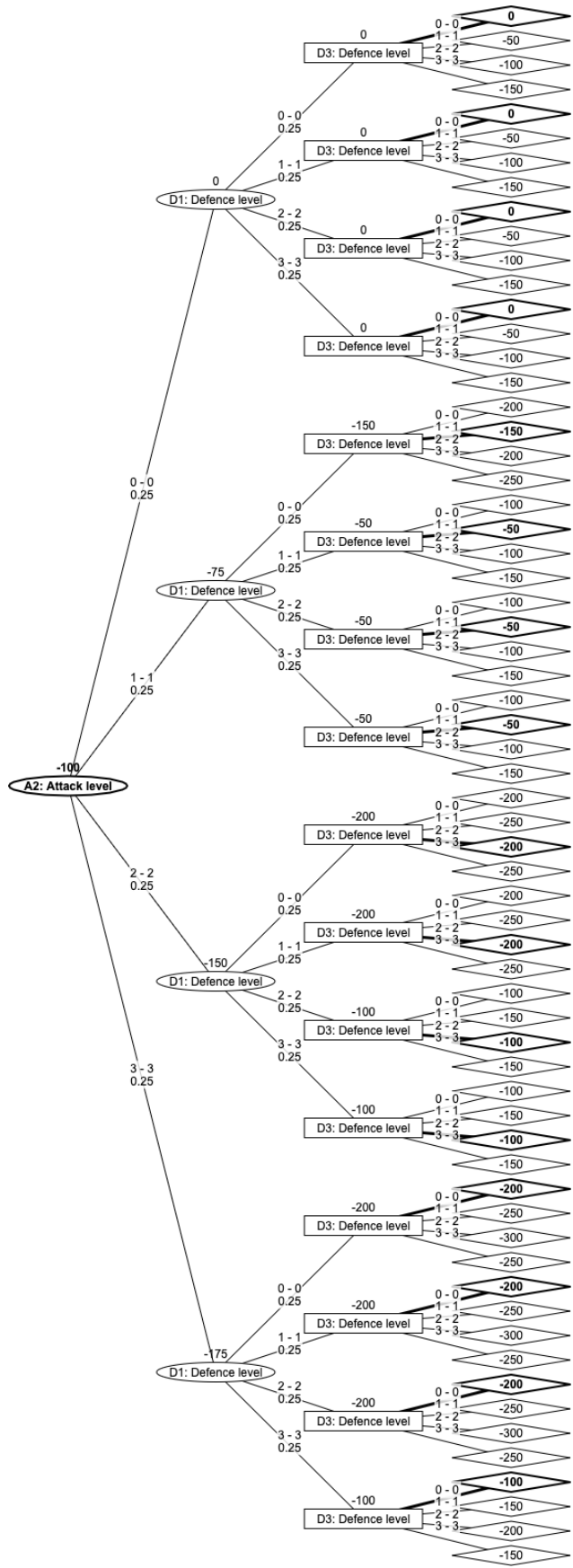

For each pair of $D_1$ and $A_2$, the defence levels at time =3 that can provide the maximum utility are the optimal defence strategies. We record the results and use them to update the CPT of the node $D_3$ as shown in Table 1.

Table 1 The updated CPT for $D_3$

| D1: ... | 0 | | | | 1 | | | | 2 | | | | 3 | | | |
|---|---|---|---|---|---|---|---|---|---|---|---|---|---|---|---|---|
| A2: ... | 0 | 1 | 2 | 3 | 0 | 1 | 2 | 3 | 0 | 1 | 2 | 3 | 0 | 1 | 2 | 3 |
| 0 | 1.0 | 0.0 | 0.5 | 1.0 | 1.0 | 0.0 | 0.5 | 1.0 | 1.0 | 0.0 | 0.5 | 1.0 | 1.0 | 0.0 | 0.5 | 1.0 |
| 1 | 0.0 | 1.0 | 0.0 | 0.0 | 0.0 | 1.0 | 0.0 | 0.0 | 0.0 | 1.0 | 0.0 | 0.0 | 0.0 | 1.0 | 0.0 | 0.0 |
| 2 | 0.0 | 0.0 | 0.5 | 0.0 | 0.0 | 0.0 | 0.5 | 0.0 | 0.0 | 0.0 | 0.5 | 0.0 | 0.0 | 0.0 | 0.5 | 0.0 |
| 3 | 0.0 | 0.0 | 0.0 | 0.0 | 0.0 | 0.0 | 0.0 | 0.0 | 0.0 | 0.0 | 0.0 | 0.0 | 0.0 | 0.0 | 0.0 | 0.0 |

The next step is to analyse the attacker's problem, which is shown in Figure 10, in the updated D-A-D model. At this step, choices of the defender are regarded as chance nodes, since $D_3$ has been determined based on $D_1$ and $A_2$, and $D_1$ is regarded as a random variable in the attacker's problem. We calculate attacker's utilities with respect to $A_2$ following formula (8). We select node $A_2$ to be the decision node, nodes $D_1$ and $D_3$ to be the chance nodes and the node A_s Utility to be the utility node in AgenaRisk and conduct influence diagram analysis. The DT representing calculation results is shown in Figure 13.

Figure 13  The DT with $A_2$ being the decision node of the attacker's problem

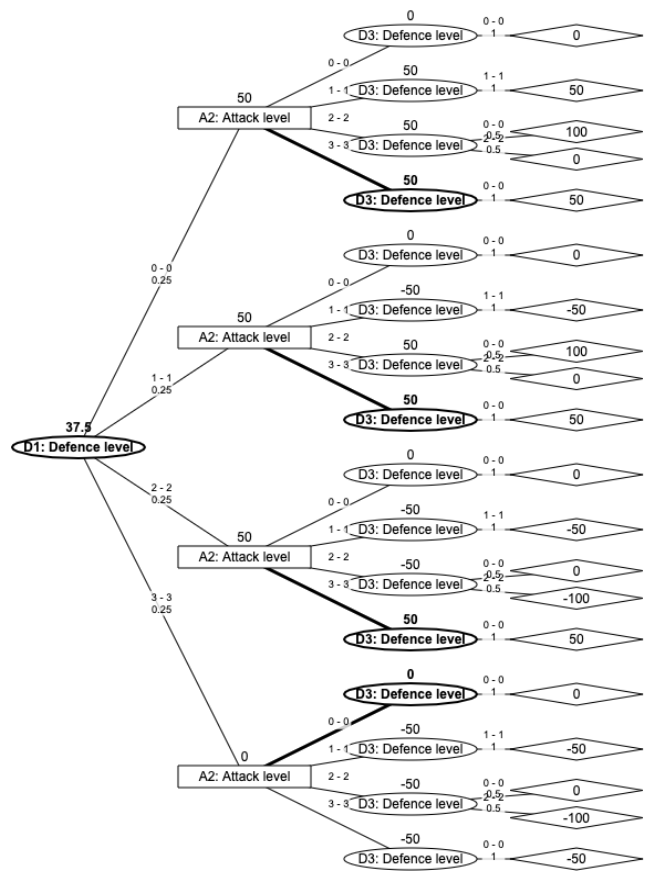

The defender can expect that the optimal choices for the attacker corresponding to each $D_1$ are attack levels which lead to maximum attacker's utilities. We record these results to update the CPT for $A_2$. The updated CPT for $A_2$ is shown in Table 2.

Table 2 The updated CPT for $A_2$

| D1:... | 0 | 1 | 2 | 3 |
|---|---|---|---|---|
| 0 | 0.0 | 0.0 | 0.0 | 1.0 |
| 1 | 0.33333334 | 0.0 | 0.0 | 0.0 |
| 2 | 0.33333334 | 0.5 | 0.0 | 0.0 |
| 3 | 0.33333334 | 0.5 | 1.0 | 0.0 |

Finally, we analyze the defender's problem with $D_1$ being the decision node in the updated D-A-D model.

The optimal choice for the defender at time t = 1 is shown by the bold path in the Figure 14 that to deploy defence level 3 and the corresponding maximum utility is 0. We have generated the optimal decisions for the defender, which are $D_1 = 3$ and $D_3 = 0$. Meanwhile, the attacker is predicted to adopt $A_2 = 0$ based on the defender's knowledge and the assumption that the attacker is intelligent and would adopt the attack that can maximize his own utility.

Figure 14  The DT with $D_1$ being the decision node in the defender's problem

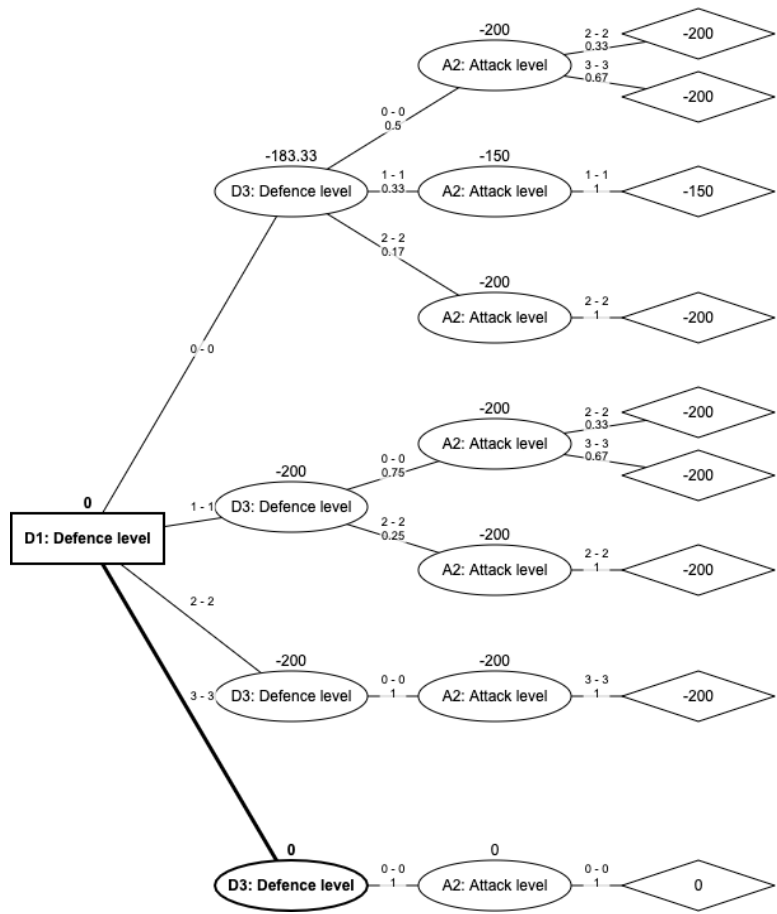

# 5. Extensions of the sequential Defend-Attack game model

In section 3, we described how to implement influence diagrams of the D-A game model using Hybrid Bayesian Networks (HBNs) and consequently how to conduct the calculation. To illustrate the calculation mechanism, we build models with core variables only, comprising decision nodes, the chance nodes representing if the attacks are successful, or not, and the utility nodes for the two agents. However, in practice, interaction between defenders and attackers may involves more factors (Banks et al., 2015). In this section, we summarize the rules required when building and calculating more complicated sequential D-A game models.

## 5.1 Rules to build and calculate the sequential Defend-Attack game models

Extending the sequential D-A game models with extra variables or longer sequences is feasible, so long as we follow these rules:

1) Decisions from $D$ and $A$ need to be made alternately. This is described as *level-k thinking* in (Banks et al., 2015). For example, a D-A-D model can be considered as the defender, the attacker, then the defender decides. In this way, (local) optimal decision for each decision phase (obtained using backwards induction) can lead to the global optimal decision set for the whole adversarial problem.

2) Each decision node is influenced by all the previous decision nodes in the sequence representing when making decision on the certain phase, the agent has the knowledge of all the previous decisions. This can be reflected by creating arcs from all its previous decision nodes $1, ..., i - 1$, for each decision node $i$, pointing to the current node $i$.

3) Set the chance nodes (i.e., success nodes) and utility nodes following dependent relations below:
   3.1) $S_i = S_i(D_{i-1}, A_i)$ $with$ $i = \{2, ..., n\}$;
   3.2) $U_D = U_D(S_2, ..., S_n, D_1, ..., D_{n-1})$;
   3.3) $U_A = U_A(S_2, ..., S_n, A_2, ..., A_n)$;

4) Decision nodes should be discrete variables to guarantee that each decision is made from finite options.

5) Decision nodes are set to follow uniform distributions to represent open mindedness when making decision.

After constructing the sequential D-A model (with extra variables or a longer sequence), we use probabilistic inference in HBNs and constructing Decision Trees (DTs) implementing the backwards induction in the sequential game model to calculate optimal strategies for the defender. In each decision phase in the sequence, we concentrate on the current decision node and regard all its previous decision nodes as chance nodes. The CPT of the current decision node is defined using probabilities conditioned on the adoption of potential strategies given all the combinations of decisions made before the current phase. The initial setting of this CPT is as a uniform distribution representing the agent's open mindedness. We can calculate the agent's utility for each decision option, conditional on each combination of previous decision, and determine the decision that provides the maximum utility value for the agent. We use this information to update the CPT of the current decision node, following the idea that, given previous decisions, the agent will only adopt decisions that lead to maximum utilities, and hence the probability of the agent adopting any decision other than this must be set to zero. Next, we move to the decision node following the current node and repeat the same operation. When we find the optimal strategy for the first decision node in the sequence, we stop and obtain the optimal strategy set for the defender and the predicted action set anticipated for the attacker with an accompanying maximum utility. We formally summarize this calculation process in Algorithm 1 below:

**Algorithm 1**: The HBN based ARA approach in solving sequential D-A models

**Initialization:**
$\Psi_A = \Psi_A(D_1, \ldots, D_{n-1}, \ldots, A_2, \ldots, A_n, V)$
$\Psi_D = \Psi_D(D_1, \ldots, D_{n-1}, \ldots, A_2, \ldots, A_n, V)$
(Assume n is even. When n is odds, the calculation follows the same process)
***for*** $(i = n, i > 0, i--)$ ***do***

    ***if*** $(i \text{ is odd})$ ***do***

        calculate:
$$a_i^* = argmax_{a_i \in A_i} \Psi_A(D_1, \ldots, d_{i+1}^*, \ldots, d_{n-1}^*, A_2, \ldots, A_i, \ldots, a_n^*, V)$$

        update the model:
$$\Psi_A = \Psi_A(D_1, \ldots, d_{i+1}^*, \ldots, d_{n-1}^*, A_2, \ldots, a_i^*, \ldots, a_n^*, V)$$
$$\Psi_D = \Psi_D(D_1, \ldots, d_{i+1}^*, \ldots, d_{n-1}^*, A_2, \ldots, a_i^*, \ldots, a_n^*, V)$$

    ***else if*** $(i \text{ is even})$ ***do***

        calculate:
$$d_i^* = argmax_{d_i \in D_i} \Psi_D(D_1, \ldots, D_i, \ldots, d_{n-1}^*, A_2, \ldots, a_{i+1}^*, \ldots, a_n^*, V)$$

        update the model:
$$\Psi_A = \Psi_A(D_1, \ldots, d_i^*, \ldots, d_{n-1}^*, A_2, \ldots, a_{i+1}^*, \ldots, a_n^*, V)$$
$$\Psi_D = \Psi_D(D_1, \ldots, d_i^*, \ldots, d_{n-1}^*, A_2, \ldots, a_{i+1}^*, \ldots, a_n^*, V)$$

***end for***
**output:**
updated model:
$$\Psi_A = \Psi_A(d_1^*, \ldots, d_{n-1}^*, a_2^*, \ldots, a_n^*, V)$$
$$\Psi_D = \Psi_D(d_1^*, \ldots, d_{n-1}^*, a_2^*, \ldots, a_n^*, V)$$
Optimal strategies for the defender:
$$\{d_1^*, \ldots, d_{n-1}^*\}$$

## 5.2 Example 1: The Defend-Attack game with extra variables

We now apply the proposed framework to a real cybersecurity problem, a simplified version of the case in (Insua et al., 2019) (Ekin et al., 2019). The problem is depicted in the ID shown in Figure 15, where the model represents a defender facing a competitor, the attacker, that may attempt a DDoS attack to undermine the availability of the defender's website, compromising her customer services and leading to a decrease in share price.

Figure 15  Influence diagram of the case study D-A problem

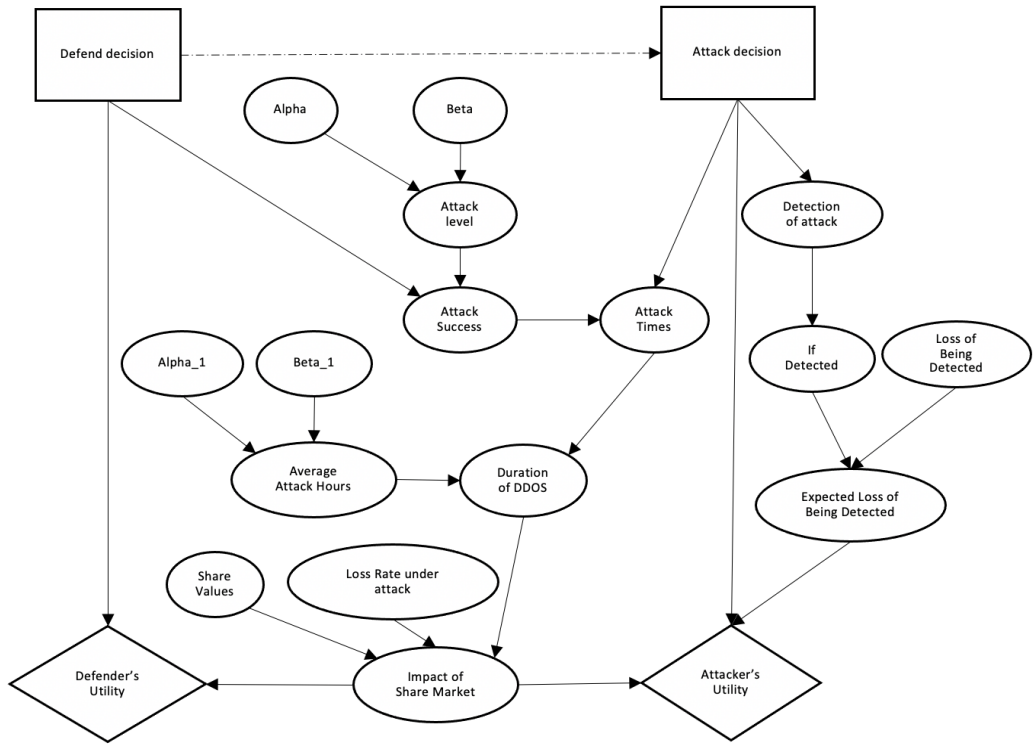

The decision node *Defend decision* ($D$) represents cloud-based DDoS (Distributed Denial of Service) protection (with states 0, 2, 5, 10 and 100 gigabits per second ($gbps$)) that the defender can deploy. The level of defence can be observed by the attacker, and therefore influences the attacker's decision, represented by the node *Attack Decision*. The node *Attack Level (AL)*, which represents the scale of the attack, is assumed to follow a Gamma distribution, with the parameters $Alpha$ and $Beta$ follow uniform distributions derived from historical data. The defence deployment $D$ and the $AL$ determine the probability of *Attack* Success ($AS$) together. The variable *Attack Times* ($AT$) is influenced by both *Attack decision* and $AS$ and is assumed to follow a Binomial distribution. The *Average Attack Hours* ($AAH$) is assumed to follow a Gamma distribution, of which the parameters $Alpha\_1$ and $Beta\_1$ also follow uniform distributions derived from historical data (Insua et al., 2019). The *Duration of DDoS* is derived from $AT$ multiplied by $AAH$. *Impact of Share Market* is derived from the organization's *Share Value* (i.e., £1500000)*, Loss Rate under Attack* and *Duration of DDoS*. In addition, the Defender's Utility ($DU$) is influenced by the Defend decision and the *Impact of Share Market*. Different defence deployment incurs different costs (i.e., 2 gbps: £2400; 5 gbps: £3600; 10 gbps: 4800; 100 gbps: £12000) (Insua et al., 2019). The defender's utility is equal to the deployment cost plus its loss in the share market.

The attacker needs to decide how many days to conduct the attack over a one-month period. This decision is represented by the node *Attack decision* with state values from 0 to 30. The longer the attack period is, the more likely the attack will be detected, represented by the node *Detection of Attack*, and is assumed to follow a Binomial distribution (Insua et al., 2019). If the attack is detected, the attacker would face legal costs, reputational costs, etc, which is represented by *Loss of being Detected*. If the attack is not detected, the *Detection Loss* will be zero. We assume the amount that the defender losses in the share market is the gain of the attacker. Based on that, *Attacker's Utility* is set to equal to *Impact of Share Market* minus *Expected Loss of Being Detected* minus the cost of *Attack Decision*.

We summarize the variables and how we assign expressions for them in AgenaRisk in Table 3.

Table 3 Variables and their expressions in Example 1

| Variable | Notation | Expression |
|---|---|---|
| Defend decision | $D$ | $Uniform\ (D), D = \{0, 2, 5, 10, 100\}$ |
| Attack decision | $A$ | $Uniform\ (0, 30)$ |
| Alpha | $a$ | $Uniform\ (4.8, 5.6)$ |
| Beta | b | $Uniform\ (0.8, 1.2)$ |
| Attack Level | $AL$ | $Gamma\ (a, b)$ |
| Attack Success | $AS$ | $min\ (max\ (AL - D, 0.0)/(D + 1.0E - 4), 1.0)$ |
| Attack Times | $AT$ | $Binomial\ (number\ of\ trials: A;\ Probability\ of\ success: AS)$ |
| Detection of Attack | $DoA$ | $Binomial\ (number\ of\ trials: A;\ Probability\ of\ success: 0.002)$ |
| If Detected | $ID$ | $if\ (DoA > 0.0, "True", "False")$ |
| Loss of Being Detected | $LBD$ | $Normal\ (Mean: 2430000, Variance: 400000)$ |
| Expected Loss of Being Detected | $ELBD$ | $\{if\ ID = False: Arithmetic\ (0.0), if\ ID = True: Arithmetic\ (LBD)\}$ |
| Alpha_1 | $a1$ | $Uniform\ (3.6, 4.8)$ |
| Beta_1 | $b1$ | $Uniform\ (0.8, 1.2)$ |
| Average Attack Hours | $AAH$ | $Gamma\ (a1, b1)$ |
| Duration of DDoS | $DoD$ | $AAH \times AT$ |
| Share Values | $SV$ | $1500000$ |
| Loss Rate under attack | $LR$ | $Uniform\ (0.00521, 0.00833)$ |
| Impact of Share Market | $ISM$ | $min\ (SV, DoD \times LR \times SV)$ |
| Defender's Utility | $DU$ | Partitional expression given status of $D$: $\{D = 0: -ISM, D = 2: -ISM - 2400, D = 5: -ISM - 3600,$ $D = 10: -ISM - 4800, D = 100: -ISM - 12000\}$ |
| Attacker's Utility | $AU$ | $ISM - ELBD - 792.0 \times A$ |

The influence diagram of this practical D-A problem with distributions of each involved variable is shown in Figure 16.

Figure 16  Influence diagram of the practical D-A problem with distributions

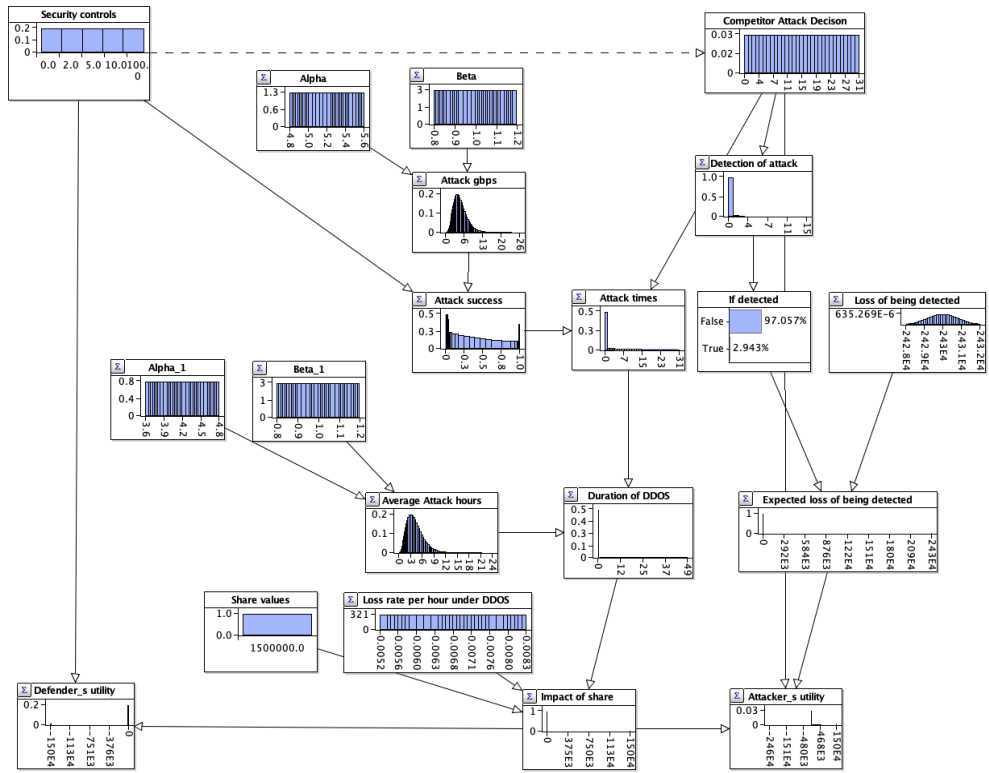

We conduct calculation using Algorithm 1 with the results as shown in Figure 17. This shows the optimal strategy for the defender is $d^* = 5$ based on her analysis of the attacker's problem, of which the optimal attack is predicted to be $a^* = 30$. In this case, the maximum utility of the defender is -3605.

Figure 17  Results of the D-A model with extra variables

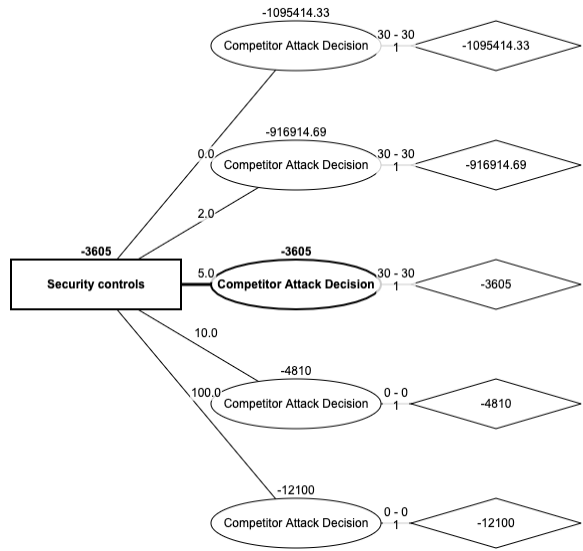

The decision tree in the process is shown in Figure 18.

Figure 18  The DT of the *Competitor Attack Decision*

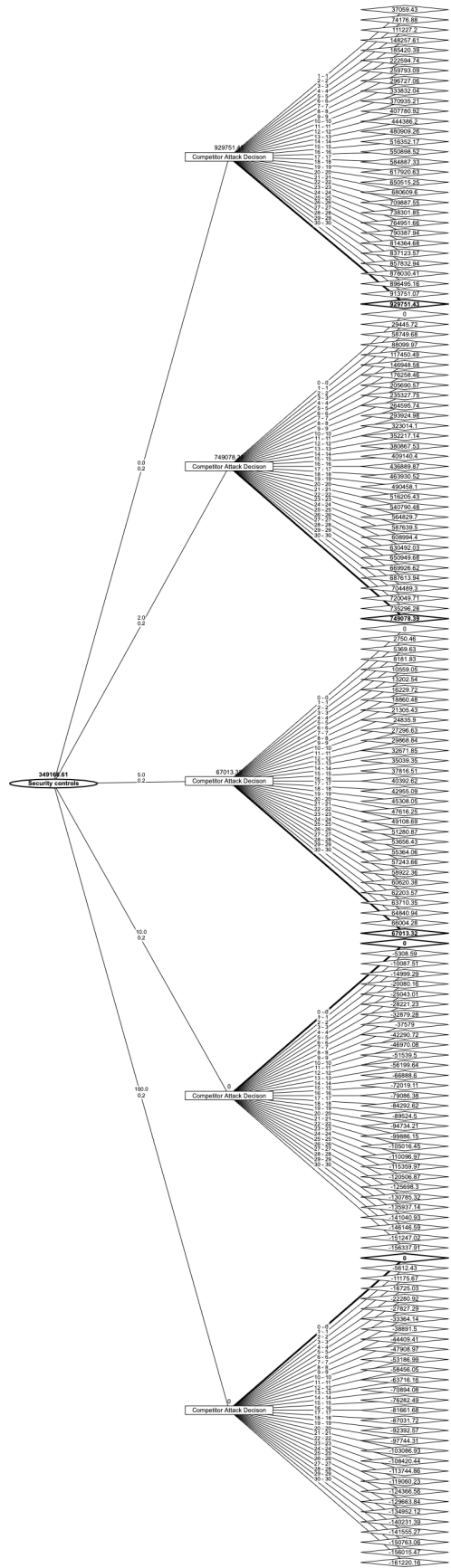

## 5.3 Example 2: The Defend-Attack game with a longer decision sequence

We now apply the proposed framework to represent and solve a practical cybersecurity problem with more rounds of interaction between the defender and the attacker. We first construct the Influence Diagram (ID) for the game using an HBN, and then illustrate how to apply Algorithm 1 to calculate the optimal strategy for the defender.

The organization provides online services for clients during the working days (Monday to Friday). Its system faces threats from a potential attacker who contemplates a DDoS attack aiming to disrupt the online service provided by the defender. To guarantee the normal operation of the online service, the defender deploys cloud-based protections against attacks. To simplify the problem, we assume the defender can adopt 0, 12, and 24 hours of protection a day, where zero hours means no protection is adopted; 12 hours means protection spreading in the whole day and the total volume is 12 hours; 24 hours is full protection. We also assume that the protection will be deployed when the defender make decision ($D_1$) on Monday and will remain valid until the next day (Tuesday). For the attacker, after observing defender's deployment, he would make an attack decision on Tuesday ($A_2$). We assume the attacker has three similar decision choices: conduct a 0, 12 or 24-hour long attack. Similarly, the attack deployment would be valid on the current day through to the next day. When the defender observes the attacker's action on Tuesday, she would make her defence decision ($D_3$) on Wednesday. This process continues until the weekend. We illustrate this adversarial problem in Figure 19.

Figure 19  The influence diagram of the Defend-Attack game with longer sequence

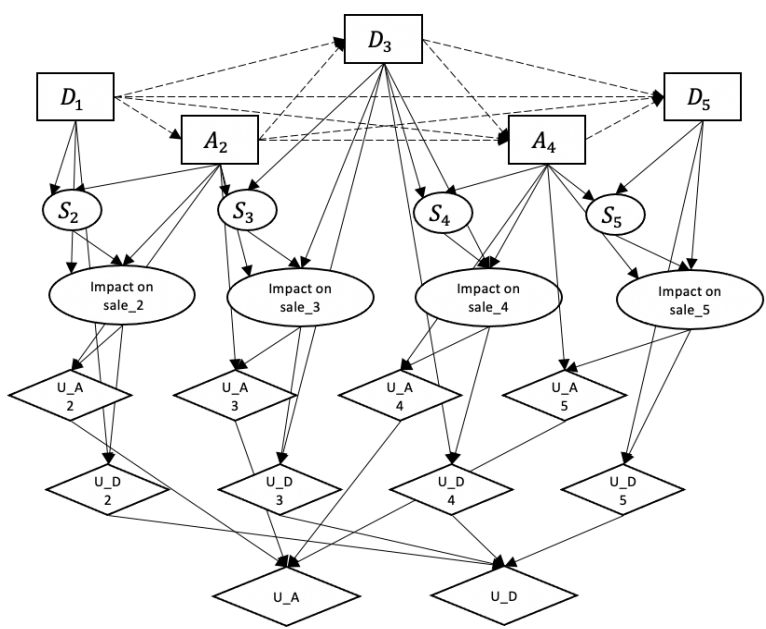

We represent decision nodes for the two agents as $D_1, A_2, D_3, A_4$ and $D_5$. For $D_1$, the node has three states (0, 12, 24) with uniform distribution. $A_2$ is conditional on $D_1$, and therefore has nine states.

Following the same rule, the decision node $D_5$ has $3^5 = 243$ states. We use $S_i$ ($i = 2, ..., 5$) to represent whether the attack is successful. In this case, we set $S_i$ to be Boolean variables that have states "true" and "false". When the attack time exceeds the defence time, some percentage of the online service will be interrupted. *Impact on sale*$_{i=2,...,5}$ represents on day $i$, how many service orders are affected by a successful attack. We calculate the percentage of unprotected hours in any day to measure those service orders interrupted by cyber-attack and assume, on average, the defender will have 1000 total online service orders and 10% of those orders may be affected by the cyber-attack. For example, on Tuesday, $A_2 = 12$, and $D_1 = 0$, then the number of interrupted orders would be $\frac{12-0}{24} \times 1000 \times 10\% = 50$. We use a TNormal distribution to represent the number of orders interrupted by the attack with uncertainty, where we set: 1) mean $= \frac{A_i - D_{i-1}}{24} \times 1000 \times 10\% = \frac{A_i - D_{i-1}}{0.24}$; 2) variance $= 400$; 3) lower bound $= 0$; 4) upper bound $= 200$. Nodes $U_{D_i}$ represent the defender's utility on day $i$, which is the cost of deploying protection (£500/hour) and the loss caused by interrupted online service (£300/order). For the attacker, we assume his utility is the gain from the organization's loss on orders minus the cost of conducting attacks (£500/hour), which on day $i$ is represented by the node $U_{A_i}$. Since utility is additive, we obtain the defender's utility over the whole week, $U_D$, from $\sum_{i=2}^{5} U\_D_i$. We do the same with the attacker's utility.

We calculate the optimal strategy using Algorithm 1 and the optimal strategy of the subgame in each decision phase can be identified using a Decision Tree representing all the possible decision path and the resulting utility. We determine the optimal decisions and use these to update the model and repeat the process until we identify the optimal strategy at the first decision node in the sequence. The results are shown in Figure 20, where the optimal strategy calculated for the defender is $\{d_1^* = 12, d_3^* = 12, d_5^* = 0\}$, while for the attacker the anticipated strategy is $\{a_2^* = 0, a_4^* = 0\}$. In this case, the maximum utility of the defender is -18 (a loss of £18,000).

Figure 20  Results of the D-A model with longer sequence

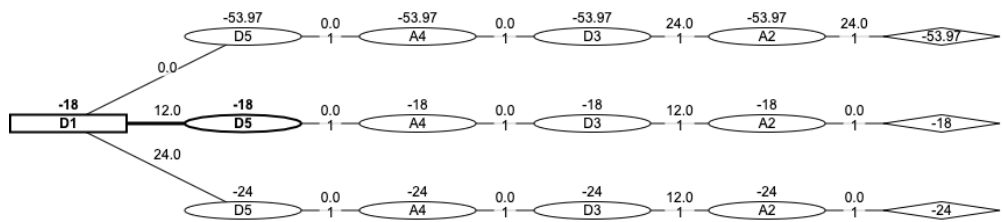

Decision trees constructed in the process are shown in Figure 21-24.

Figure 21　The DT of D5

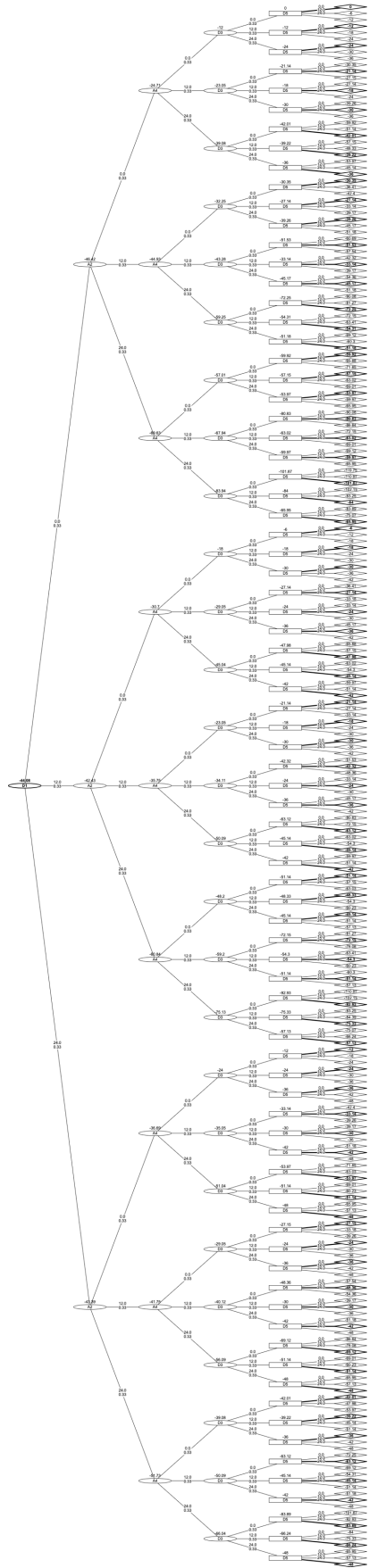

Figure 22  The DT of A4

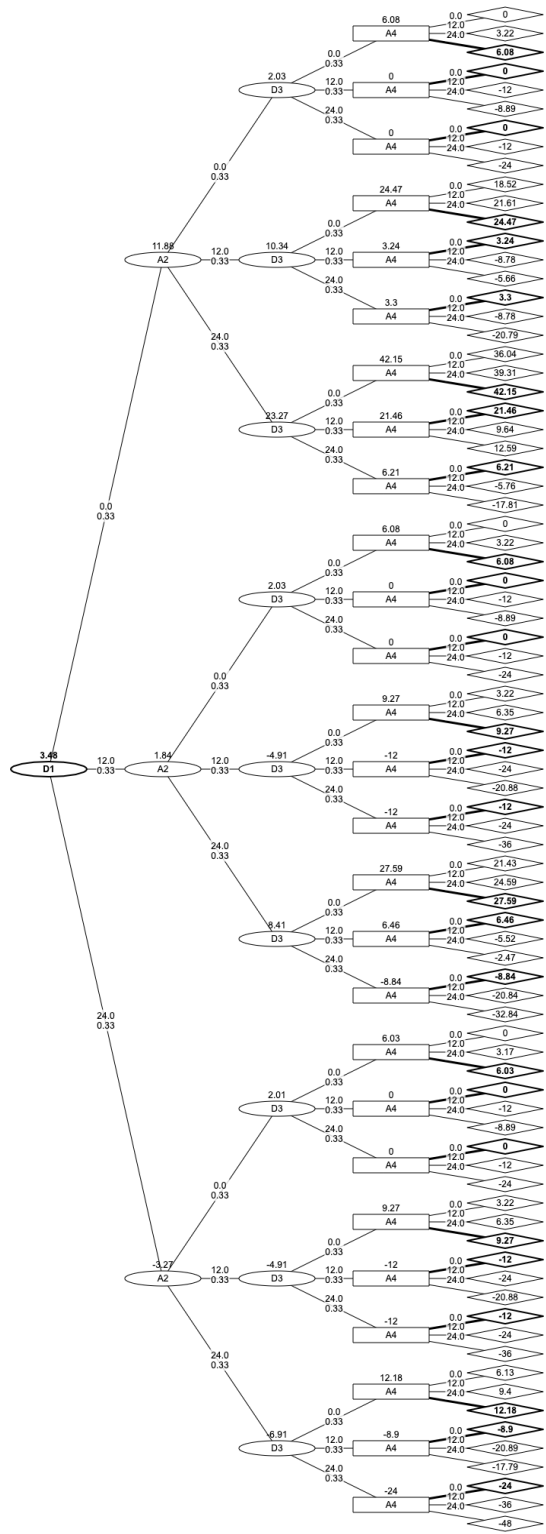

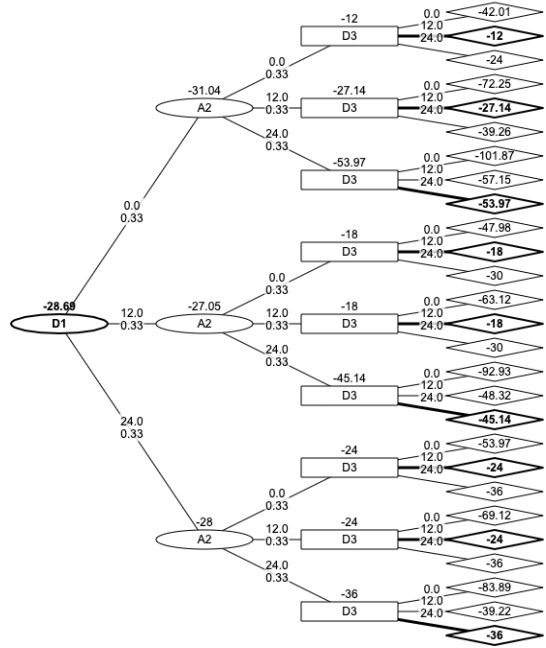

Figure 24  The DT of A2

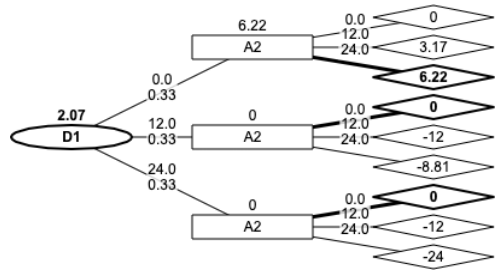

# 6. Supporting the defender's dynamic decision making

In a game with a longer decision sequence, $k$-level thinking (i.e., he anticipates, she anticipates, what he would anticipates) is involved. As is shown in example 2, the defender anticipates what the attacker would do based on her belief of the attacker's utility (assuming that implementing an attack and a defence costs the same per hour and the loss of the defender caused by attack is the gain of the attacker) and the attack success probability. Based on the analysis of the attacker's problem, the defender determines her optimal decision at the first place. However, the original decision-making problem can become a dynamic decision-making problem over time, meaning that the defender can always use the fresh-observed data to update the model and make real-time-updated decision accordingly. More precisely, in example 2, the calculated strategies for the defender are optimal standing on Monday without observing any other information. When it comes to Wednesday, the defender can make decision based on what she can observe, i.e., the attack that actually taken or its consequences on Tuesday, and what she can anticipates, i.e., the attack that likely to be conducted in the future (on Thursday). Hence,

Dynamic Decision Analysis (DDA), which involves updating the D-A game model using observable information through the decision sequence, is required.

In this section, we provide the DDA algorithm to deal with dynamic decision making. Moreover, we represent two examples to illustrate how the proposed DDA algorithm can be applied to analyse practical problems modelled by HBN.

## 6.1 The algorithm for dynamic decision analysis

The core idea of supporting dynamic decision making here is to update the sequential D-A model with real-time data and conduct decision analysis on the model updated using this data. We consider this dynamic decision-making issue based on example 2. In the example, we have obtained the optimal strategy set for the defender standing on Monday. This strategy set includes the optimal strategies suggested for the defender on Monday ($D_1$), Wednesday ($D_3$) and Friday ($D_5$) based on her prediction of attacks on Tuesday and Thursday ($A_2$ and $A_4$ respectively). Hence, initially, the defender would take the optimum decision $d_1^*$ on Monday. On Wednesday, the defender can then observe effects of the adopted action or its consequence from the previous day, Tuesday in this case. If the actual attack is observable, we use $a_2$ to represent the observed attack (which does necessarily need to equal to the predicted attack $a_2^*$). Then we would use the calculated $d_1^*$ and the observed $a_2$ (rather than the calculated $a_2^*$) to update the sequential D-A model; otherwise, if only the attack consequence is observable (i.e., whether the attack succeeded on Tuesday), we use this new observation for $S_2$ to update the model. In the latter case, if the attack on Tuesday is not observable, we remove the arc pointing from node $A_2$ to $D_3$. In addition, to represent the fact that $D_3$ is influenced by $S_2$, we add an arc pointing from $S_2$ to $D_3$. Then we conduct dynamic decision analysis based on the updated model.

We formally summarize the process of supporting dynamic decision-making using Algorithm 2.

**Algorithm 2**: The HBN based ARA approach in solving dynamic decision making

---

**Initialization:**

$\Psi_A = \Psi_A(D_1, \dots, D_{n-1}, \dots, A_2, \dots, A_n, V)$

$\Psi_D = \Psi_D(D_1, \dots, D_{n-1}, \dots, A_2, \dots, A_n, V)$

(Assume n is even. When n is odds, the calculation follows the same process)

***for*** $(j = 1, j \leq n - 1, j = j + 2)$ ***do***

    In the latest model conduct ***Algorithm 1*** and get **outputs:**

        The updated model:

$$\Psi_A = \Psi_A(d_1^*, \dots, d_{n-1}^*, a_2^*, \dots, a_n^*, V)$$

$$\Psi_D = \Psi_D(d_1^*, \dots, d_{n-1}^*, a_2^*, \dots, a_n^*, V)$$

        Optimal strategy set of the game:

$$\{d_1^*, \dots, d_j^*, \dots, d_{n-1}^*, a_2^*, \dots, a_{j+1}^*, \dots, a_n^*\}$$

    identify optima for $D_j$, which is $d_j^*$;

    ***if*** (the actual adopted attack is observable)

    ***then do***

        record the actual attack adopted in day $j + 1$, which is $A_{j+1} = a_{j+1}$;

        update the model using $D_j = d_j^*$ and $A_{j+1} = a_{j+1}$;

    ***else if*** (only the consequence of attack is observable)

    ***then do***

        record the consequence of attack in day $j + 1$, which is $S_{j+1} = s_{j+1}$;

        update the model using $D_j = d_j^*$ and $S_{j+1} = s_{j+1}$;

        remove arcs: $A_{j+1} \text{-->} D_{j+2}, \dots, D_{n-1}$

        add the arc: $S_{j+1} \text{-->} D_{j+2}$

    ***end if***

    recover decision nodes in day $j + 2, j + 3, \dots, n$ to be uniform.

    updated model:

$$\Psi_A = \Psi_A(d_1^*, \dots, d_j^*, \dots, D_{n-1}, \dots, a_2, \dots, a_{j+1}^*, \dots, A_n, V)$$

$$\Psi_D = \Psi_D(d_1^*, \dots, d_j^*, \dots, D_{n-1}, \dots, a_2, \dots, a_{j+1}^*, \dots, A_n, V)$$

***end for***

**output:**

Updated model:

$$\Psi_A = \Psi_A(d_1^*, \dots, d_{n-1}^*, a_2, \dots, a_n, s_2, \dots, s_n, V)$$

$$\Psi_D = \Psi_D(d_1^*, \dots, d_{n-1}^*, a_2, \dots, a_n, s_2, \dots, s_n, V)$$

Optimal strategies for the defender:

$$\{d_1^*, \dots, d_{n-1}^*\}$$

---

## 6.2 Example 3: the actual attacks are observable

In this subsection, we show how we can apply algorithm 2 to support dynamic decision-making based on example 2 with the actual attacks are observable. According to the results calculated in Example 2, $\{d_1^* = 12, a_2^* = 0, d_3^* = 12, a_4^* = 0, d_5^* = 0\}$, the defender would deploy $D_1 = d_1^* = 12$ on Monday as her optimal move. However, when it comes to Wednesday and the defender is going to deploy the defence, we assume she realizes that the attacker attacks i.e., 24 hours on Tuesday rather than 0 hour represented by $a_2^* = 0$. The strategy for this is to use $D_1 = d_1^* = 12$ and $A_2 = a_2 = 24$ updating the model. We enter observations of $D_1$ and $A_2$ into the model shown in Figure 9. The updated HBN representing the defender's problem and the attacker's problem on Wednesday is shown in Figure 25, while the ID showing distributions of variables are illustrated in Figure 26.

Figure 25  The updated HBNs for the DDM on Wednesday.

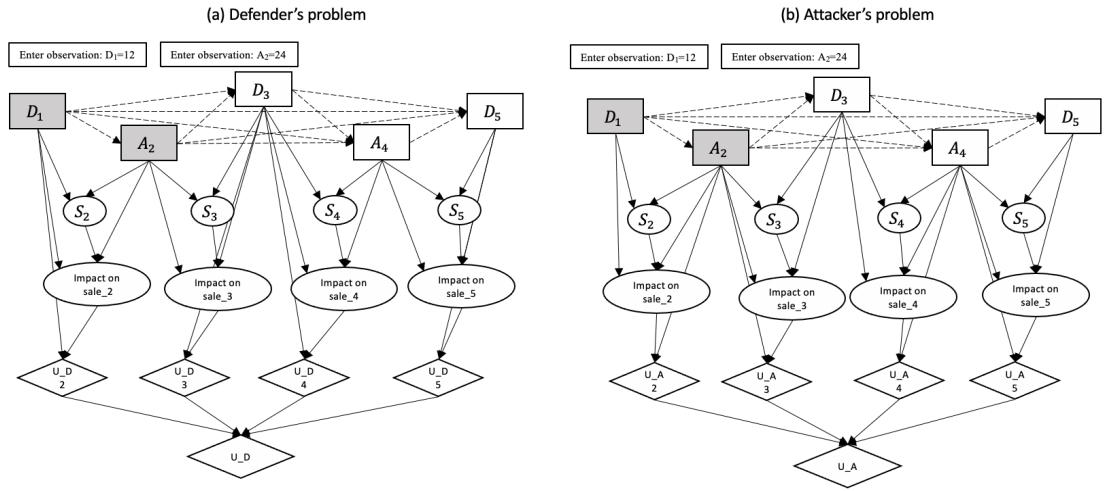

Figure 26  The updated HBNs for the DDM on Wednesday-with distribution of variables.

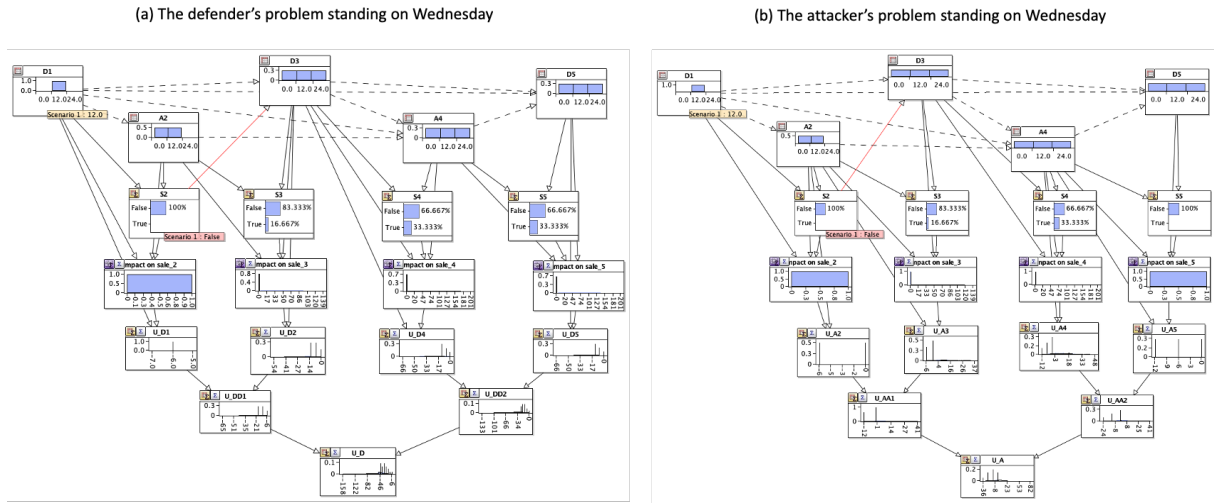

Then to calculate the optimal strategy for the defender standing on Wednesday, we apply Algorithm 2 to the updated HBNs. The optimal strategy for the defender on Wednesday is represented by the DT in Figure 27. Hence, we get $d_3^* = 24$.

Figure 27 The DT of D3 given information observed before Wednesday

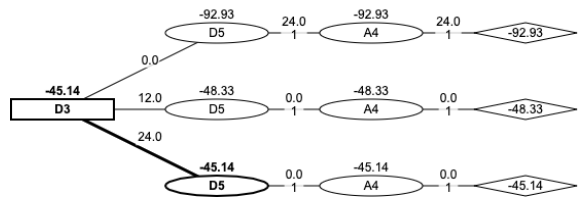

Decision trees constructed in the process are shown in Figure 28 and 29.

Figure 28 The DT of D5 given information observed on Wednesday

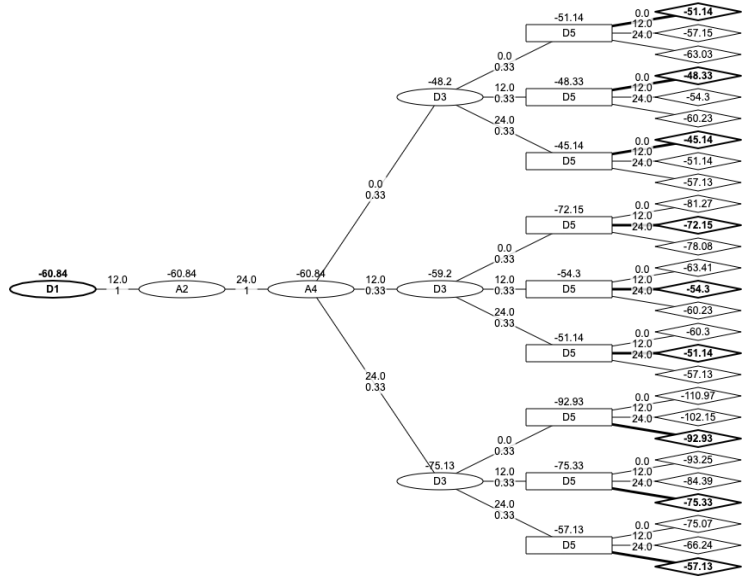

Figure 29 The DT of A4 given information observed on Wednesday

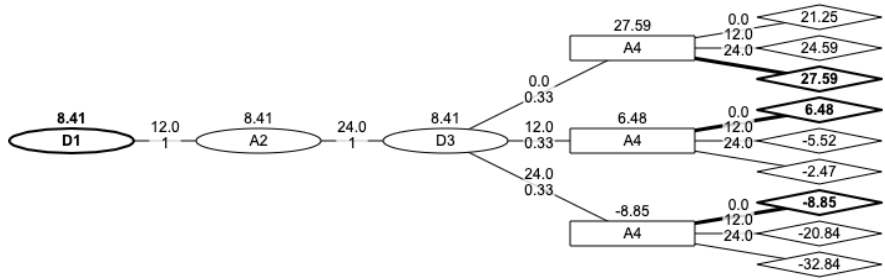

Finally, when it comes to Friday, given observed attack on Thursday, we can calculate the defender's optimal choice following the similar way. Assuming that the attacker eventually adopted $A_4 = a_4 = 12$ on Thursday, we update the model by entering the observations $D_3 = d_3^* = 24$ and $A_4 = a_4 = 12$. Then we construct a DT of $D_5$ to determine the best choice for the defender when she makes the defence decision on Friday. We show this DT in Figure 30, which represents that $D_5 = d_5^* = 12$ is the optimal choice when the defender stands on Friday.

Figure 30 The DT of D5 given information observed on Friday

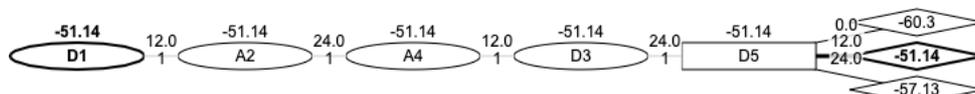

Hence, we support the defender with optimal decisions in the dynamic process that deploy $D_1 = 12$ at the first place, deploy $D_3 = 24$ when observing the attack on Tuesday is $A_2 = 24$ and deploy $D_5 = 12$ when observing the attack on Thursday is $A_4 = 12$. We illustrate observations, predictions and the optimal strategy set of this dynamic decision-making process in Figure 31.

Figure 31  Results summary of dynamic decision making in Example 3

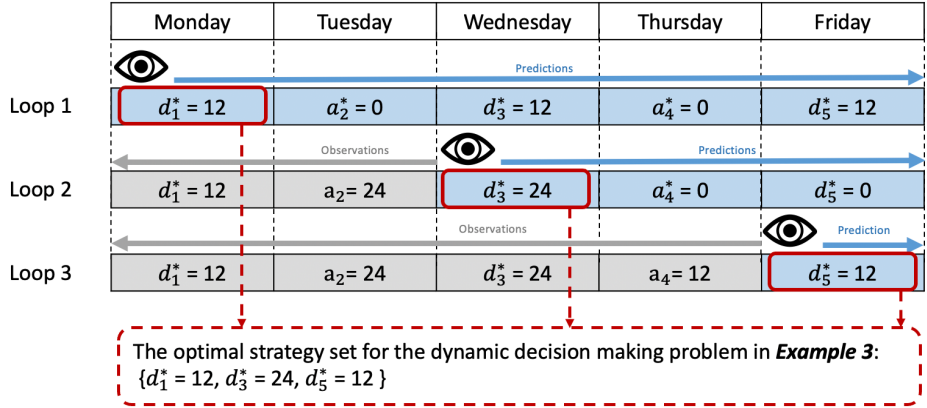

## 6.3 Example 4: only the consequences of attacks are observable

In this subsection, we illustrate how we apply algorithm 2 to support dynamic decision-making based on example 2 with only the consequences of attacks are observable.

Let's start from considering how to determine the optimal decision for the defender when it comes to Wednesday. At this time point, the defender has conducted $D_1 = d_1^* = 12$ on Monday and can observe if the attack succeeded on Tuesday (represented by the node $S_2$). We remove the arc pointing from node $A_2$ to $D_3$, since under the assumption of this example, the past attack is no more observable. In addition, to represent $D_3$ is influenced by $S_2$, we add an arc pointing from $S_2$ to $D_3$. Then we update the model using the real-time data which includes $D_1 = d_1^* = 12$ and the observed states of $S_2$ (i.e., assuming $S_2$ = False). The updated model is shown in Figure 32 and 33. Algorithm 2 can be then applied to the updated HBNs, where $D_1$ and $S_2$ are chance nodes while $D_3$, $A_4$ and $D_5$ are decision nodes. The corresponding optimal strategy set can be then calculated.

Figure 32  The updated HBNs for the DDM problem on Wednesday.

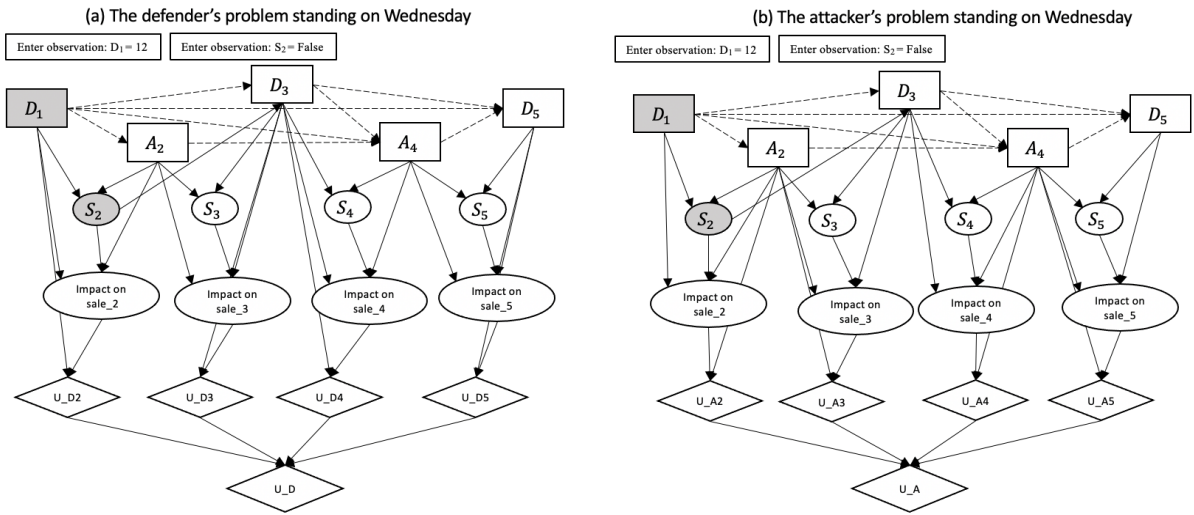

Figure 33  The updated HBNs for the DDM problem on Wednesday-with distribution of variables.

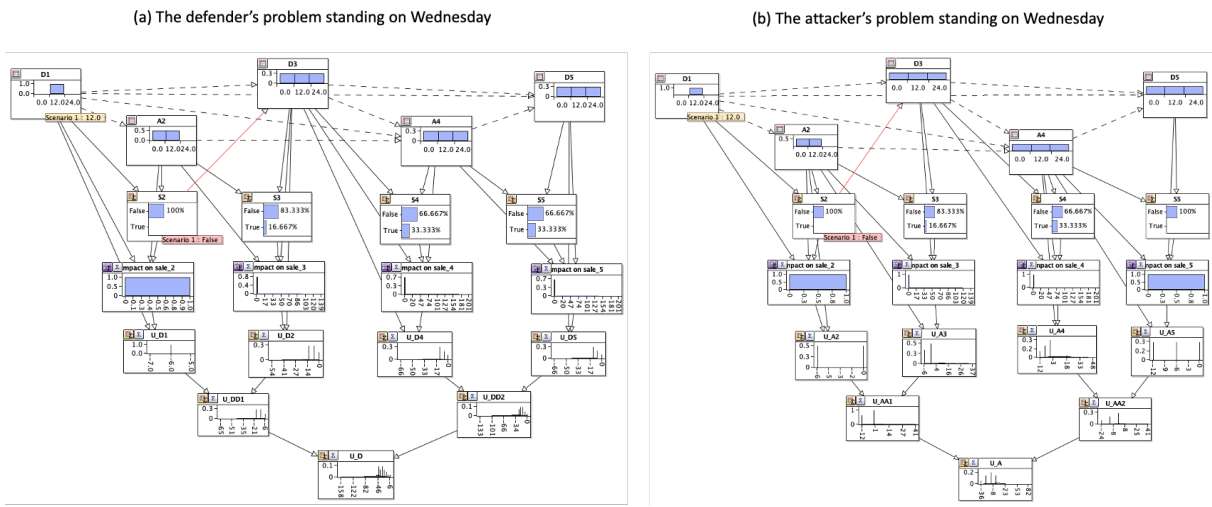

By constructing the DT for the defender on Wednesday, which is represented in Figure 34, we know that the optimal decision for the defender standing on the current time point (Wednesday) is $d_3^* = 12$. The optimal strategy set for all the decision nodes at this stage is $\{d_3^* = 12, a_4^* = 0, d_5^* = 0\}$. DTs constructed in the process are shown in Figure 35 and 36.

Figure 34  The DT of D3 given information observed on Wednesday

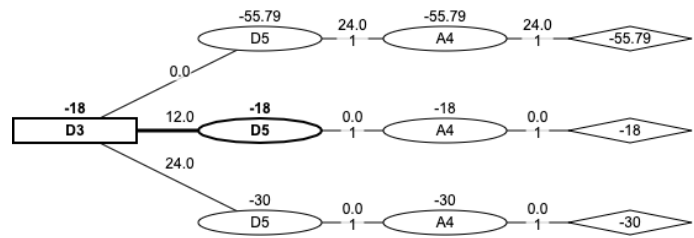

Figure 35  The DT of D5 given information observed on Wednesday

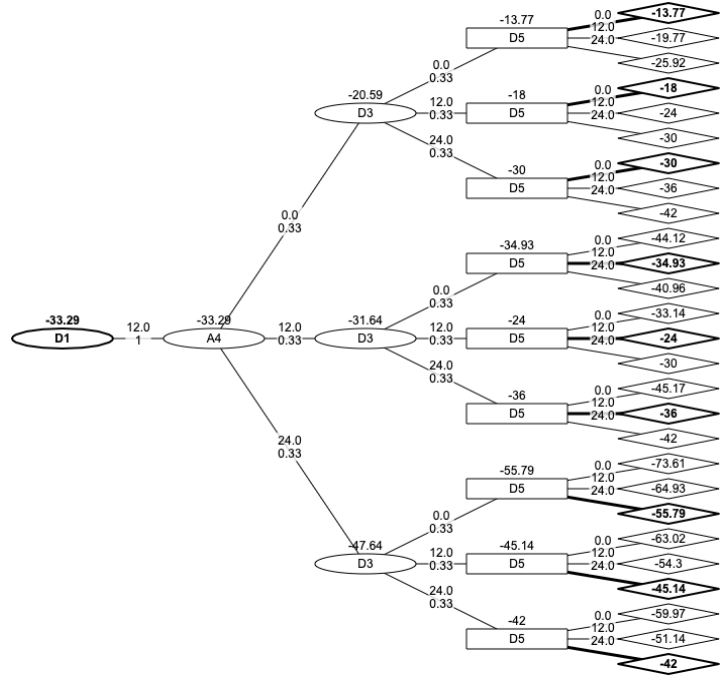

Figure 36  The DT of A4 given information observed on Wednesday

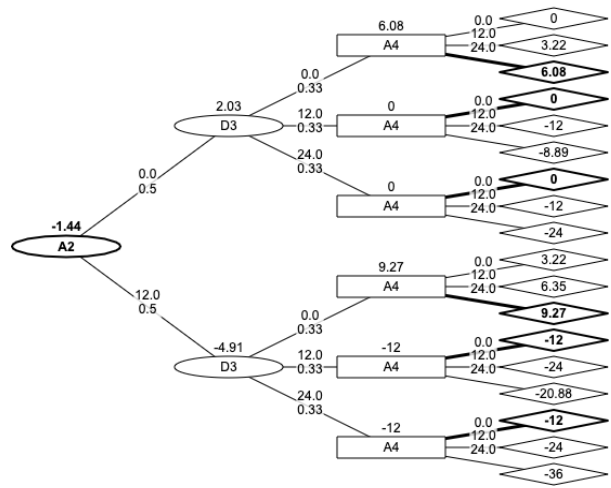

The newly calculated strategy $d_3^*$ is the optimal one for the defender at the current time point (Wednesday) given observed information have been considered to update her mind while she still uses her best knowledge to predict what attack would be adopted on Thursday.

The defender would then adopt this optimal strategy on Wednesday. When it comes to Friday, she needs to update her mind again with the newly observed information and furthermore determines the best move on Friday. We can assume she observes that the attack on Thursday succeeded. Since the attack on Thursday is unobservable in reality, we remove the arc pointing from node $A_4$ to $D_5$. Meanwhile, to represent $D_5$ is influenced by $S_4$, we add an arc pointing from $S_4$ to $D_5$. In this stage, the decision node is only $D_5$ and there are no attack decisions needed to be predicted for determining the optimal $D_5$. The updated HBN is shown in Figure 37 and 38.

Figure 37 The updated HBN for the defender's problem on Friday.

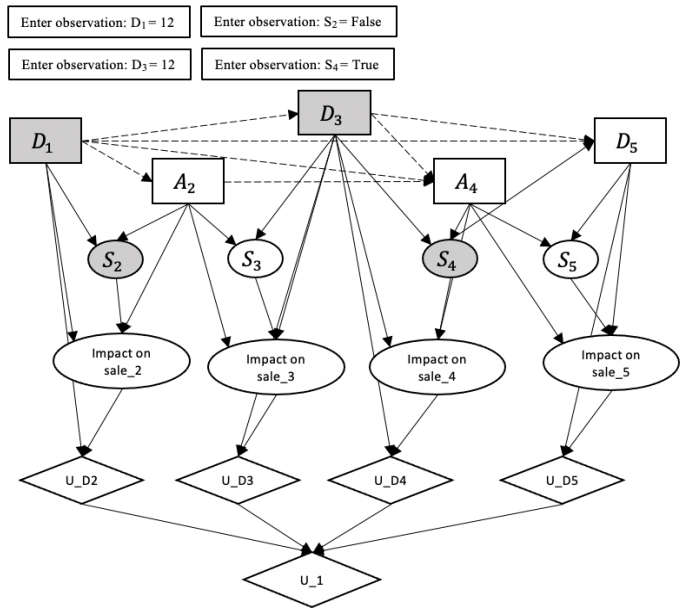

Figure 38 The updated HBN for the defender's problem on Friday-with distribution of variables.

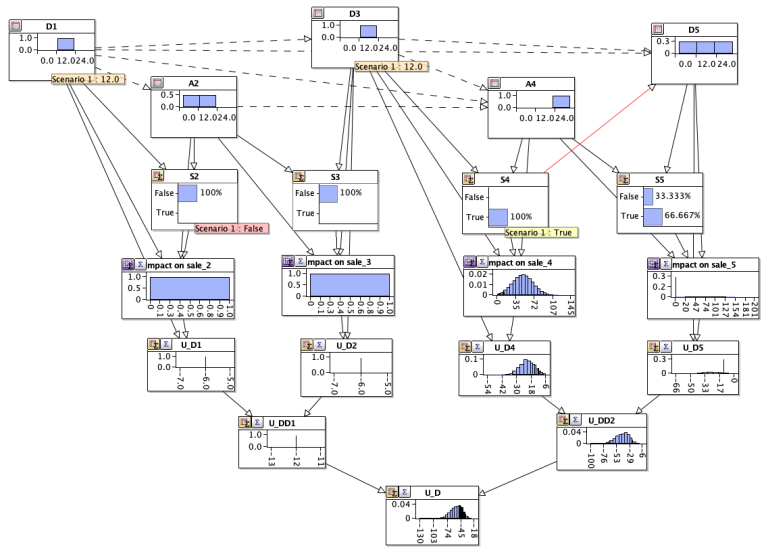

Based on the updated model, we construct the corresponding DT which is shown in Figure 39. It can be represented that the optimal strategy for the defender standing on Friday is $d_5^* = 24$.

Figure 39 The DT of D5 given information observed on Friday

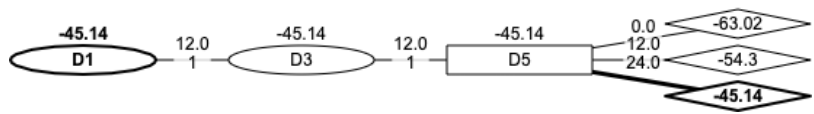

Hence, we support the defender with optimal decisions in the dynamic process that deploy $D_1 = 12$ at the first place, deploy $D_3 = 12$ when observing the attack on Tuesday fails and deploy $D_5 = 12$ when observing the attack on Thursday successes. We illustrate observations, predictions and the optimal strategy set of this dynamic decision-making process in Figure 40.

Figure 40  Results summary of dynamic decision making in Example 4

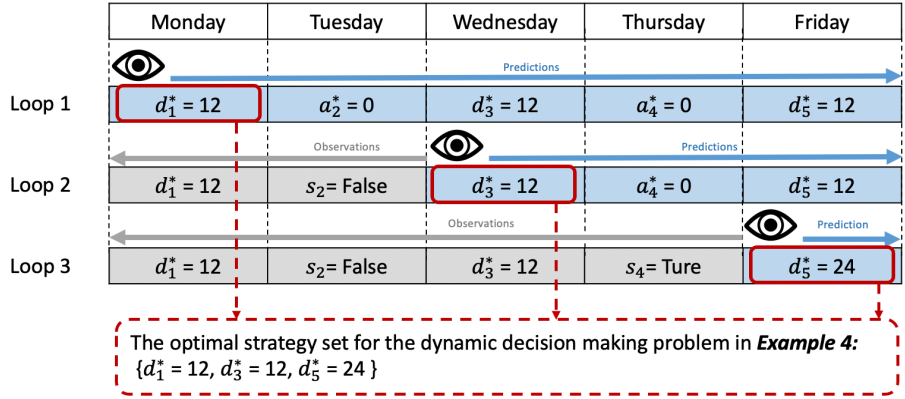

## 7. Conclusion

We propose an HBN based ARA approach for supporting decision making in sequential defend-attack game problems. This kind of problem is typically extracted to the sequential D-A model. We illustrate how to use the proposed method to calculate the optimal strategy in this template. Furthermore, to model more complicated cases that may be likely in practice, we consider two extended sequential D-A templates involving extra variables and longer decision sequence respectively. We construct the algorithm based on HBNs and the ARA approach to calculate optimal decisions for the supported agent (the defender) and provide examples to illustrate how the proposed method can be applied. Since the applied HBN inference provides an automated way to compute hybrid D-A models and extends their use to involve mixtures of continuous and discrete variables, the proposed HBN based ARA approach is more versatile compared with the Monte Carlo (MC) based ARA approach. More importantly, the proposed approach is novel in that it supports dynamic decision making whereby new real-time observations can be employed to update the D-A model timely and optimal decisions can be determined based on both generic information from the past and rigorous anticipation about the future. This dynamic decision analysis mechanism can make more effective use of information and can better simulate the actual decision-making process. Examples are provided, illustrating how the proposed framework can be adjusted for decision analysis in more complicated but more practical scenarios and serving as template for further expansion according to practical application requirement.

## Acknowledgements

Jiali Wang is supported by a China Scholarship Council (CSC)/Queen Mary Joint PhD scholarship. Agena Ltd provided the AgenaRisk software gratis.

# Appendix A: Notations summary

| Notation | Definition and Explanation |
|---|---|
| ARA | Adversarial Risk Analysis |
| D-A model | Defend-Attack model |
| HBN | Hybrid Bayesian Network |
| MC simulation | Monte Carlo simulation |
| IDs | Influence Diagrams |
| HIDs | Hybrid Influence Diagrams |
| DTs | Decision Trees |
| CPT | Conditional Probability Table |
| $D = \{d_1, \ldots, d_m\}$ | A set of defences that the defender can choose from. |
| $A = \{a_1, \ldots a_n\}$ | A set of attacks that the attacker can choose from. |
| $U_D(D, A, S)$ | The utility of the defender given D, A and whether the attack successes (S). |
| $\psi_D(d, a)$ ($\psi_A(d, a)$ is defined in the similar way) | The expected utility that the defender obtains when the decisions are $(d, a) \in D \times A$; $$\psi_D(d, a) = P_D(S = 0\|d, a) \times U_D(d, a, S = 0) \\ + P_D(S = 1\|d, a) \times U_D(d, a, S = 1)$$ |
| $P_D(S\| d, a)$ | The probability of successful attack given the decisions are $(d, a)$. |
| $a^*(d)$ | $= argmax_{a \in A} \psi_A(d, a), \forall d \in D$; The optimal strategy in A that can maximize $\psi_A(d, a)$. |
| $d^*$ | $= argmax_{d \in D} \psi_D(d, \ a^*(d))$; The optimal strategy in D that can maximize $\psi_D(d, a)$ given $a = a^*(d)$. |